\documentstyle[psfig]{mn}
\begin{document}

\def\reef{\par\noindent\hang}
\def\etal{et al.\ }
\def\eg{{\em eg.\ }}
\def\etc{{\em etc.\ }}
\def\ie{{\em i.e.\ }}

\def\spose#1{\hbox to 0pt{#1\hss}}
\def\approxlt{\mathrel{\spose{\lower 3pt\hbox{$\sim$}}
	\raise 2.0pt\hbox{$<$}}}
\def\approxgt{\mathrel{\spose{\lower 3pt\hbox{$\sim$}}
	\raise 2.0pt\hbox{$>$}}}
	
\def\Mdot{\hbox{$\dot M$}}
\def\degmark{$^\circ$}
\def\<{\thinspace}
\def\s{\hbox{\phantom{5}}}	%one space
\def\ss{\s\s}		%two spaces
\def\sss{\ss\s}		%three
\def\ssss{\ss\ss}	%four
\def\lit{\obeyspaces\obeylines}
%
% 	Simple units
\def\arc{{\rm\thinspace arcsec}}
\def\cm{{\rm\thinspace cm}}
\def\ct{{\rm\thinspace ct}}
\def\erg{{\rm\thinspace erg}}
\def\eV{{\rm\thinspace eV}}
\def\g{{\rm\thinspace g}}
\def\G{{\rm\thinspace G}}
\def\ga{{\rm\thinspace gauss}}
\def\K{{\rm\thinspace K}}
\def\keV{{\rm\thinspace keV}}
\def\m{{\rm\thinspace m}}
\def\km{{\rm\thinspace km}}
\def\kpc{{\rm\thinspace kpc}}
\def\Lsun{\hbox{$\rm\thinspace L_{\odot}$}}
\def\rad{{\rm\thinspace rad}}
\def\MeV{{\rm\thinspace MeV}}
\def\Mpc{{\rm\thinspace Mpc}}
\def\Msun{\hbox{$\rm\thinspace M_{\odot}$}}
\def\pc{{\rm\thinspace pc}}
\def\ph{{\rm\thinspace photons}}
\def\s{{\rm\thinspace s}}
\def\yr{{\rm\thinspace yr}}
\def\sr{{\rm\thinspace sr}}
\def\Hz{{\rm\thinspace Hz}}
\def\GHz{{\rm\thinspace GHz}}
\def\W{{\rm\thinspace W}}
%	Compound units
\def\cmps{\hbox{$\cm\s^{-1}\,$}}
\def\ctps{\hbox{$\ct\s^{-1}\,$}}
\def\cmsq{\hbox{$\cm^2\,$}}
\def\cmcu{\hbox{$\cm^3\,$}}
\def\pHz{\hbox{$\Hz^{-1}\,$}}
\def\pcmcu{\hbox{$\cm^{-3}\,$}}
\def\ergcmcups{\hbox{$\erg\cm^3\ps\,$}}
\def\ergpcmps{\hbox{$\erg\cm^{-3}\s^{-1}\,$}}
\def\ergpcmsqps{\hbox{$\erg\cm^{-2}\s^{-1}\,$}}
\def\ergpspcmsq{\hbox{$\erg\cm^{-2}\s^{-1}\,$}}
\def\ergpspkpcsq{\hbox{$\erg\s^{-1}\kpc^{-2}\,$}}
\def\ergpspA{\hbox{$\erg\s^{-1}\AA^{-1}\,$}}
\def\ergpcmsqpspsqarcsec{\hbox{$\erg\cm^{-2}\s^{-1}\arc^{-2}\,$}}
\def\ergpcmsqpspapsqarcsec{\hbox{$\erg\cm^{-2}\s^{-1}\AA^{-1},\arc^{-2}\,$}}
\def\ergps{\hbox{$\erg\s^{-1}\,$}}
\def\gpcm{\hbox{$\g\cm^{-3}\,$}}
\def\gpcmps{\hbox{$\g\cm^{-3}\s^{-1}\,$}}
\def\gps{\hbox{$\g\s^{-1}\,$}}
\def\WpHz{\hbox{$\W\Hz^{-1}\,$}}
\def\kmps{\hbox{$\km\s^{-1}\,$}}
\def\ksec{\hbox{$ksec\,$}}
\def\Lsunppc{\hbox{$\Lsun\pc^{-3}\,$}}
\def\Msunpc{\hbox{$\Msun\pc^{-3}\,$}}
\def\Msunpkpc{\hbox{$\Msun\kpc^{-1}\,$}}
\def\Msunppc{\hbox{$\Msun\pc^{-3}\,$}}
\def\Msunppcpyr{\hbox{$\Msun\pc^{-3}\yr^{-1}\,$}}
\def\Msunpyr{\hbox{$\Msun\yr^{-1}\,$}}
\def\pcm{\hbox{$\cm^{-1}\,$}}
\def\pcmsq{\hbox{$\cm^{-2}\,$}}
\def\pmsq{\hbox{$\m^{-2}\,$}}
\def\radpmsq{\hbox{$\rad\m^{-2}\,$}}
\def\pcmcuK{\hbox{$\cm^{-3}\K$}}
\def\phps{\hbox{$\ph\s^{-1}\,$}}
\def\phpcmsqps{\hbox{$\ph\cm^{-2}\s^{-1}\,$}}
\def\pHz{\hbox{$\Hz^{-1}\,$}}
\def\pMpc{\hbox{$\Mpc^{-1}\,$}}
\def\pMpccu{\hbox{$\Mpc^{-3}\,$}}
\def\ps{\hbox{$\s^{-1}\,$}}
\def\psqcm{\hbox{$\cm^{-2}\,$}}
\def\psr{\hbox{$\sr^{-1}\,$}}
\def\pyr{\hbox{$\yr^{-1}\,$}}
\def\kmpspMpc{\hbox{$\kmps\Mpc^{-1}$}}
\def\Msunpyrpkpc{\hbox{$\Msunpyr\kpc^{-1}$}}
\def\Hb{H$\beta$\ }
\def\Hbn{H$\beta$}
\def\otw{[OII]$\lambda$3727\ }
\def\oth{[OIII]$\lambda$5007\ }
\def\otwn{[OII]$\lambda$3727}
\def\othn{[OIII]$\lambda$5007}

\title{Optical integral field spectroscopy of the extended line emission
around six radio-loud quasars}

\author[C.S. Crawford and C. Vanderriest]{C.S. Crawford$^{1,\dag}$ \& 
C. Vanderriest$^{2}$\\
1. Institute of Astronomy, Madingley Road, Cambridge CB3 0HA \\
2. Observatoire de Paris-Meudon (DAEC), 92195 Meudon Cedex, France \\
$^{\dag}$ Visiting Astronomer, Canada-France-Hawaii Telescope, which is
operated by the National Research Council of Canada,\\
 Centre National de Recherche Scientifique, and the University of Hawaii.}

\maketitle

\begin{abstract}

\noindent
We present integral field spectroscopy of a small sample of radio-loud
quasars at intermediate redshift ($0.26<z<0.60$), most of which are
associated with large radio sources. All have oxygen line emission
extended over tens of kiloparcsecs, and these nebulae display a
diverse range in both morphology and kinematic behaviour. Two quasars
show \lq plumes' of extended line emission, two show a clumpy
structure and a further one shows a smooth distribution. There is no
clear pattern with regard to the distribution of the ionized gas in
relation to the radio source axis; the extended emission-line regions
are found both parallel and perpendicular -- and also totally
unrelated to - the radio axis. The velocity structure of the ionized
gas ranges from essentially static to apparent smooth rotation, and in
two cases, show a clear association with the radio source. Given the
disparity in properties, the nebulae all show a surprisingly similar
ionization state, as measured by the extended lines of \otw and \othn.
Assuming the gas is ionized by at least the nearby quasar nucleus, we
use the emission line ratios to infer a pressure in the ionized gas;
in all cases we find it to be at high pressure, suggesting confinement
by an external (probably intracluster) medium.\\

%\noindent {\em Note that Figs and Tabs are correctly numbered, but do not
%necessarily appear in the right order in this file .}

\end{abstract}


\section{Introduction}
% common place occurrence for RL objects at range of z
Spatially extended emission-line regions are commonplace around
steep-spectrum radio-loud quasars; at low (Stockton \& MacKenty 1987,
hereafter SM87; Crawford, Fabian \& Johnstone 1988; Hutchings \&
Crampton 1990; Durret \etal 1994; Boisson \etal 1994), intermediate (Crawford \&
Fabian 1989; Forbes \etal 1990; Hutchings 1992; Bremer \etal 1992;
Ridgway \& Stockton 1998) and high redshift (Heckman \etal 1991a,b).
The gas emits strongly in the coolants of Ly$\alpha$, \otw and
\oth over radii of tens to a hundred \kpc\ from the quasar nucleus,
thus providing a relatively direct way to study of the immediate
environment of a quasar beyond its host galaxy.
% spatial distirbution
The spatial distribution of the ionized gas appears highly structured
into clumps and filaments, but does not appear to be well correlated
with any extended continuum structure (SM87). Despite the statistical
association of the presence of an EELR with steep-spectrum radio
emission, few objects at low redshift show any clear morphological
relation of the extended emission-line region (EELR) to the extended
radio source structure (SM87; Hes, Barthel \& Fosbury 1996). An
alignment of the principal axes is seen more obviously at at earlier
epochs (Heckman \etal 1991a,b; Crawford \& Vanderriest 1997; Ridgway
\& Stockton 1998).

% mergers
The distribution of the ionized gas and extended continuum
(particularly when elongated into putative tidal tails) is often taken
as indicative of the emitting gas -- and probably the quasar nuclear
activity -- originating in a gravitational interaction between the quasar
host galaxy and a nearby companion (SM87; Hutchings \& Neff 1992;
Stockton \& Ridgway 1991; Durret \etal 1994; Chatzichristou,
Vanderriest \& Jaffe 1999).
% kinematic structure
The kinematic structure of the ionized gas, however, where mapped
shows no obvious organized pattern, being chaotic and sometimes
split into several velocity components (Hickson \& Hutchings 1987;
Durret \etal 1994).

% ionization structure and implications and relation to cluster environment
Off-nuclear spectroscopy of the extended oxygen emission shows the
 emitting gas to be of surprisingly low ionization given its proximity
 to a luminous quasar nucleus (Crawford \& Fabian 1989; Forbes \etal
 1990; Bremer \etal 1992; Boisson \etal 1994; Heckman etal 1991a;
 Crawford \& Vanderriest 1997). The pressure of the gas is deduced to
 be high, typically $kT>4\times10^6$\pcmcuK\ within 20\kpc\ of the
 quasar; if unconfined, this will disperse rapidly on less than a
 crossing time.  To avoid this -- given the frequent occurrence of
 such systems and their apparently chaotic velocity structure --
 either an exceptionally large reservoir of neutral gas is required,
 or the EELR must be confined within a high-pressure environment.The
 inferred pressure is consistent with that in the hot intracluster
 medium of groups and clusters of galaxies (eg Fabian 1994 and
 references therein). There is plentiful additional evidence from
 other wavebands to support the idea that powerful intermediate-(and
 higher-)redshift radio sources lie in environments richer than the
 field: from excess galaxy counts (Yee \& Green 1984, 1987; Yates et
 al 1989; Hill \& Lilly 1991; Ellingson, Yee \& Green 1991; Dickinson
 1997), host continuum profiles that resemble cD galaxies (Best et al
 1998) and gravitational lensing signatures (Bower \& Smail 1997;
 Deltorn et al 1997), to asymmetric Faraday radio depolarization
 (Garrington \& Conway 1991), sources with very high Faraday rotation
 measures (Carilli \etal 1994, 1997) and direct X-ray detection of the
 environment (Crawford \& Fabian 1993; Worrall \etal 1994; Crawford \&
 Fabian 1995, 1996a,b; Crawford \etal 1999; Dickinson \etal 1999;
 Hardcastle \& Worrall 1999). A surrounding intracluster medium could
 possibly provide fuel for the nuclear activity from a
 cooling flow gravitationally focussed to the quasar (Fabian \&
 Crawford 1990).

% similarity to RGs EELRS
%Many of the properties of the EELR associated with $z>0.6$ radio-loud
%quasars are similar to those around powerful radio galaxies at similar
%redshift (Heckman \etal 1991a,b; Crawford \& Vanderriest 1997; Ridgway
%\& Stockton 1997). This strongly supports schemes uniting the
%appearance of FR~II radio sources according to their orientation with
%respect to the viewer (eg Barthel 1989).

% rationale for IFS, previous work on this
Much of the previous work on quasar EELRs has had to rely on either
narrow-band imaging of the whole structure, or spectroscopy of only a
small region of the gas. In this paper we present optical area
spectroscopy of a small sample of low redshift ($0.2\le z\le0.6$) radio-loud
quasars in order to gain a complete picture of the ionized gas by
combining direct information on its distribution, ionization state and
kinematics over the whole nebula. Connections and comparisons can then
be made both directly between these properties and those of the
associated radio source.

Only a few low- to intermediate-redshift quasars have been previously targeted
using integral field spectroscopy. Durret \etal (1994) presented
optical area spectroscopy of three low-redshift quasars, two of which
are in common with the sample presented here. The EELR were found to
have a clumpy distribution, with discrete blobs embedded in a more
diffuse envelope surrounding the quasar. The velocity fields were
chaotic, not simply fit by a single rotating disk model. 
%The
%wavelength range observed included only the (redshifted) \Hb and \oth
%emission lines, and so a full picture of the ionization state could
%not be included in the analysis (narrow \Hb is not so strong in the
%EELR).
We have presented area spectroscopy data of the exceptional line emission
associated with the $z=0.734$ radio-loud quasar 3C254 (Crawford \&
Vanderriest 1997).  The ARGUS instrument used (Vanderriest 1995;
Chatzichristou etal 1999) allows a larger aperture and a wider wavelength
range to be used than the TIGRE system in Durret \etal (1994). The
kinematics and distribution of the large EELR around 3C254 strongly indicate
an interaction of a radio jet with a dense cloud of ionized gas; however
this appears to have little effect on the ionization state of the gas, which
remains low in this region.

The layout of this paper is as follows: we detail the observations and
data reduction in section 2; section 3 presents a brief introduction to 
and our results on each of the six quasars in order of increasing
redshift; we summarize and discuss our findings in
section 4; finally, conclusions are presented in section 5.

% motivation?
%origin of this gas? its role in the AGN activity? 
%is its visibility affected by the radio source? 
%similarity to EELR around rgals?

\section{ Observations}

\subsection{Sample selection}
We selected our targets from those radio-loud quasars already known to
possess spatially extended emission-line regions at the time of the
observations, and with a redshift $z<0.8$ such that both the dominant
emission lines of \otw and \oth could be observed. In addition, we chose
quasars with a range of radio source properties -- from fairly compact to
very extended -- from those thought to lie in groups and clusters of
galaxies. Additional considerations included whether an X-ray flux from the
quasar was known at the time, enabling a better knowledge of the
photo-ionizing continuum from the nucleus.

\subsection{Observations and Data reduction}

The targets were observed using the integral field spectroscopy device
ARGUS on the Canada-France-Hawaii Telescope during the nights of 1993
June 24-27, except for the observation of 3C215 which was observed on
the night of 1995 Jan 7. ARGUS is an additional mode to the MOS-SIS
spectrograph on the CFHT (Vanderriest 1995) which uses a flexible
bundle of optical fibres to transform independent spectra collected at
an $\sim8.5\times12.4$ arcsec hexagonal aperture in the Cassegrain
focal plane, into a \lq pseudo' long-slit for collimation through the
MOS. Each fibre has a diameter of 0.4 arcsec, and the spectral range
and resolution of the observation are determined only by the CCD and
grism used in MOS, the same for all the field of view. The weather
conditions were photometric, with seeing conditions typically just
sub-arcsecond. The total exposure on each quasar was obtained by
co-adding consecutive observations of 1500-2500\s\ each; a log of
observations is given in Table~\ref{tab:obslog}. The Loral~3 CCD was
used with the O300 grism, leading to a wavelength coverage of
3980--9680\AA\ (the combined CCD and grism efficiency is effectively
zero below 3980\AA) at a dispersion of 3\AA\ per pixel and a spectral
resolution of $\sim11$\AA. The ARGUS aperture was oriented with north
to the top of the aperture for all the observations in June~1993, and
with north toward the top left-hand corner of the aperture for the
observation of 3C215 in Jan~1995.

\begin{table*}
\caption{Log of the observations. \label{tab:obslog} }
\begin{tabular}{lllllllcc}
         &      &          &        &         &          &       & & \\
         & Name & Redshift & Date   & Airmass & Exposure & A$_V$ & Seeing & position angle  \\
         &      &          &        &         & (sec)    &       &
(arcsec) & (degrees) \\
\hline
0903+169 & 3C215   & 0.412 & Jan 95 & 1.212  & 7000  & 0.240  & 0.8 & -45 \\
1305+069 & 3C281   & 0.599 & Jun 93 & 1.346  & 7500  & 0.143 & 0.8 & 0\\
1512+370 & 4C37.43 & 0.371 & Jun 93 & 1.316  & 4500  & 0.092 & 0.8 & 0\\
1545+210 & 3C323.1 & 0.266 & Jun 93 & 1.102  & 4000  & 0.287 & 1.0 & 0\\
1618+177 & 3C334   & 0.555 & Jun 93 & 1.144  & 5400  & 0.274 & 1.0 & 0\\
2251+113 & 4C11.72 & 0.323 & Jun 93 & 1.036  & 6500  & 0.336 & 1.1 & 0\\
%1623+269 & 4C26.48 & 0.779 & Jun 93 & 1.138  & 5000  & 0.073 & 1.0 & 0 \\
         &      &          &        &         &          &       & & \\
\end{tabular}
\newline
%\noindent *=light leak affecting row 5 \\
\end{table*}

The data were reduced in IRAF using the steps described in detail for the
quasar 3C254 in Crawford \& Vanderriest (1996). In summary, after correction
for bad CCD columns, the individual frames of each quasar were
median-combined to remove cosmic ray events and then bias-subtracted (with a
bias estimated from the zero response at the blue-wavelength part of the chip).
The data were corrected for spatial distortion, flat-fielded using a
normalised exposure of a tungsten lamp, and wavelength-calibrated using
exposures of a Neon-Helium lamp. They were then flux-calibrated, corrected
for both atmospheric extinction, and for Galactic reddening along the line
of sight (by the $A_V$ listed in Table~\ref{tab:obslog}, which was estimated
from Galactic hydrogen column densities in the direction of each quasar
;Stark \etal 1992), and converted using the relation of Bohlin, Savage \&
Drake (1978), assuming R of 3.2 and the reddening law of Cardelli, Clayton
\& Mathis (1989)). The data were then separated into spectra from individual
fibres.  Sky (and scattered-light) subtraction used the average spectrum
from fibres that were neither too near the edge of the aperture, nor too
close to the source itself; typically a total of 5 or 6 sky fibres per row.
A separate sky spectrum was constructed to subtract from each row of the
hexagon, using the mean spectrum of all chosen sky fibres in that and the
two adjacent rows.

Like the data presented in  Crawford \& Vanderriest (1996; 1997), the flux calibration is
imperfect in that some flux has been lost from the calibration data at
wavelengths greater than 7000\AA. In practice this has little consequence
for the observations presented here, as we are modelling the dynamics,
distribution and ionization state of the extended emission-line region
(EELR) using the [OII]$\lambda3727$, \Hb and [OIII]$\lambda$5007
emission lines. Only two of our sample quasars (3C281 and 3C334) are
at high enough redshift for some of the lines to be observed at wavelengths
appreciably beyond 7000\AA. For these two quasars the flux calibration
error has been simply corrected by comparing the well-defined
power-law slope of the quasar nuclear continuum to high-quality
optical/UV spectra in the literature (c.f. CV96). Even if this
correction is not sufficient, only widely-spaced line intensity ratios
(\eg [OII]/[OIII]) may be affected in these two quasars -- the
morphology of the EELR, its dynamics, and its ionization state as
deduced from the [OIII]/\Hb intensity ratio are not affected. (The
error in flux calibration is the same all over the aperture, CV96.)

Arc-lamp exposures were used to characterize fibre-to-fibre variations in
the spectral output across the aperture, at the wavelengths of redshifted
[OIII] (and [OII], if observed) in each quasar. Variations across the
aperture include a decrease in the instrumental resolution toward the
edge of the aperture in the blue, and also slight systematic shifts in
line wavelength between fibres (CV96). The results of the arcline fitting
are used to remove these systematics from the emission-line results
presented in this paper

\subsection{Emission-line fitting}

The spectral fitting of the emission lines was done using QDP (Tennant
1991). Small complexes of neighbouring lines were fit together (eg narrow
and broad \Hb with [OIII]$\lambda\lambda$4959,5007) over a wavelength range
of a few hundred \AA ngstroms in each individual fibre spectrum. Narrow lines
within such a complex were constrained to have the same redshift and velocity
width as each other, and were fit by a Gaussian (a satisfactory fit even to
the [OII] doublet at this resolution). Where an emission line was unresolved
(generally only toward the edges of the ARGUS aperture), the FWHM used in
the fitting was fixed at an average of the immediately neighbouring fibres'
line fits.  

\section{Results}

We present an introduction to and results for individual quasars in order of
increasing redshift.

%(6.5 arcmin from the centre) 
% nb this is a regular cluster with its own cd. 3c323.1 is _not_ in the cd!!
\subsection {3C323.1, $z$=0.266}
3C323.1 is one of the nearest radio-loud quasars, and is located on
the outskirts of the compact cluster of galaxies Z1545.1+2104 (Oemler,
Gunn \& Oke 1972; Hintzen \& Scott 1978; Yee \& Green 1984).  It is
associated with a steep-spectrum triple radio source which is straight
and symmetric over a $\sim360$\kpc\ diameter (Bogers \etal 1994). The
quasar lies in a luminous elliptical host galaxy (Neugebauer, Matthews
\& Armus 1995; Bahcall \etal 1997), with several continuum companions.
The dominant companion is a compact galaxy located 2.7 arcsec
approximately West at a similar redshift to the quasar
(Neugebauer
\etal 1995; Canalizo \& Stockton 1997); there are two other objects in
the field, one at 19 arcsec East and one further North (eg McLeod
 \& Rieke 1994; Hes \etal 1996).  Hutchings, Johnson \&
Pyke (1988) find continuum
condensations 1.3 arcsec NW and 3 arcsec S, after subtraction of the quasar
light. 3C323.1 has long been known to show an asymmetric emission-line
region, extended approximately from south-east to west across the quasar
core  (SM87; Hes \etal 1996).

\begin{figure}
%\vspace{-1.0cm}
% 3c323.1/results/newfit.idlcom
% alternatively, the better colour version is in /colour/3c323.1col.ps
%\psfig{figure=fig1.ps}
\psfig{figure=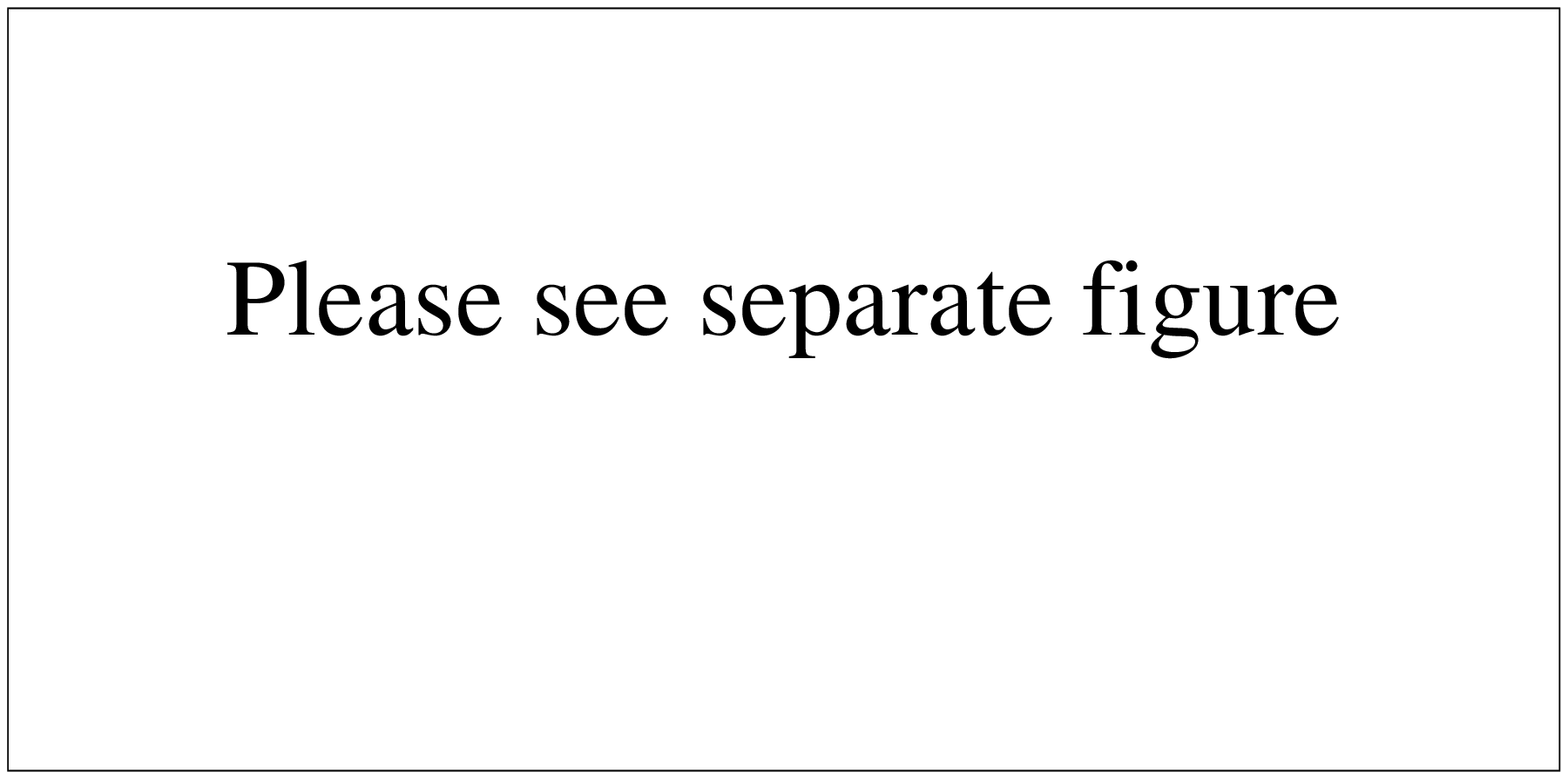,width=0.45\textwidth,angle=270}
%\vspace{-2.0cm}
\caption{\label{fig:3c323.1im}
Reconstructed ARGUS images of 3C323.1 in \oth intensity (top left; plotted
on a scale of 0 (white) to a maximum (black) of
$4.45\times10^{-15}$\ergpspcmsq, the lowest detection being
0.03$\times10^{-15}$\ergpspcmsq); in radial velocity of the \oth line
relative to the nucleus (middle right; on a scale from $-360$ to 250 \kmps
as shown in the colour bar); and in broad \Hb intensity showing the
approximate PSF of the observation (lower left; plotted on a scale of
$0-11.28\times10^{-15}$\ergpspcmsq with a minimum detection of
0.12$\times10^{-15}$\ergpspcmsq). The position of the peak fibre is marked
by a cross in the velocity image, and the orientation of the radio source
axis is marked by a thick line. The position of the edge of the hexagonal
ARGUS aperture is marked around each image. North is to the top and east to
the left, and 2 arcsec corresponds to $\sim10$\kpc\ at the redshift of the
quasar.}
\end{figure}

The spectral shape of the broad \Hb emission line is slightly skew (eg Brotherton
 1996)  and thus not perfectly fit by a symmetric gaussian
centred about the narrow line component; otherwise fitting models to
[OIII]+\Hb in QDP is straightforward. Whilst the ARGUS aperture encompasses
the companion galaxy to the west of the quasar core, the continuum detection
level of our observation is not sufficiently sensitive to detect it (the efficiency of the
ARGUS system is only around half that of a long-slit spectrum). We detect
the continuum light from the host galaxy of the quasar as   marginally
more extended than the (combined MOS+ARGUS) instrumental point-spread
function (PSF), similar to the (slightly asymmetric) distribution of the
quasar in seen in broad \Hbn. We present reconstructed images of 3C323.1 in
the light of broad \Hb and [OIII] in Fig~\ref{fig:3c323.1im}: a disc is
drawn at the position of each fibre where line emission was significantly
detected -- areas left blank within the outline of the ARGUS aperture have
no line detection in those fibres. The greyscale colour within each disk gives the intensity of
the line emission measured within that fibre's spectrum.
Fig~\ref{fig:3c323.1im} shows the [OIII] line emission to be clearly more
extended than the approximate PSF given by the broad \Hb image. The nebula
extends out to a maximum of 4.1 arcsec (21\kpc) to the East (at p.a.
105\degmark), and to $\sim1.9$ arcsec ($\sim$10kpc) in all other
directions, with the maximum dimension approximately perpendicular to
the radio source axis.  (1 arcsec corresponds to $\sim5.1$\kpc\ at the redshift of the
quasar, assuming the cosmology of $H_0$=50\kmpspMpc and $q_0$=0.5 which will
be used throughout this paper).

\begin{figure}
%\vspace{-1.0cm}
% /data/csc/argus/argus/jundata/3c323.1/results/diag.idlcom
\psfig{figure=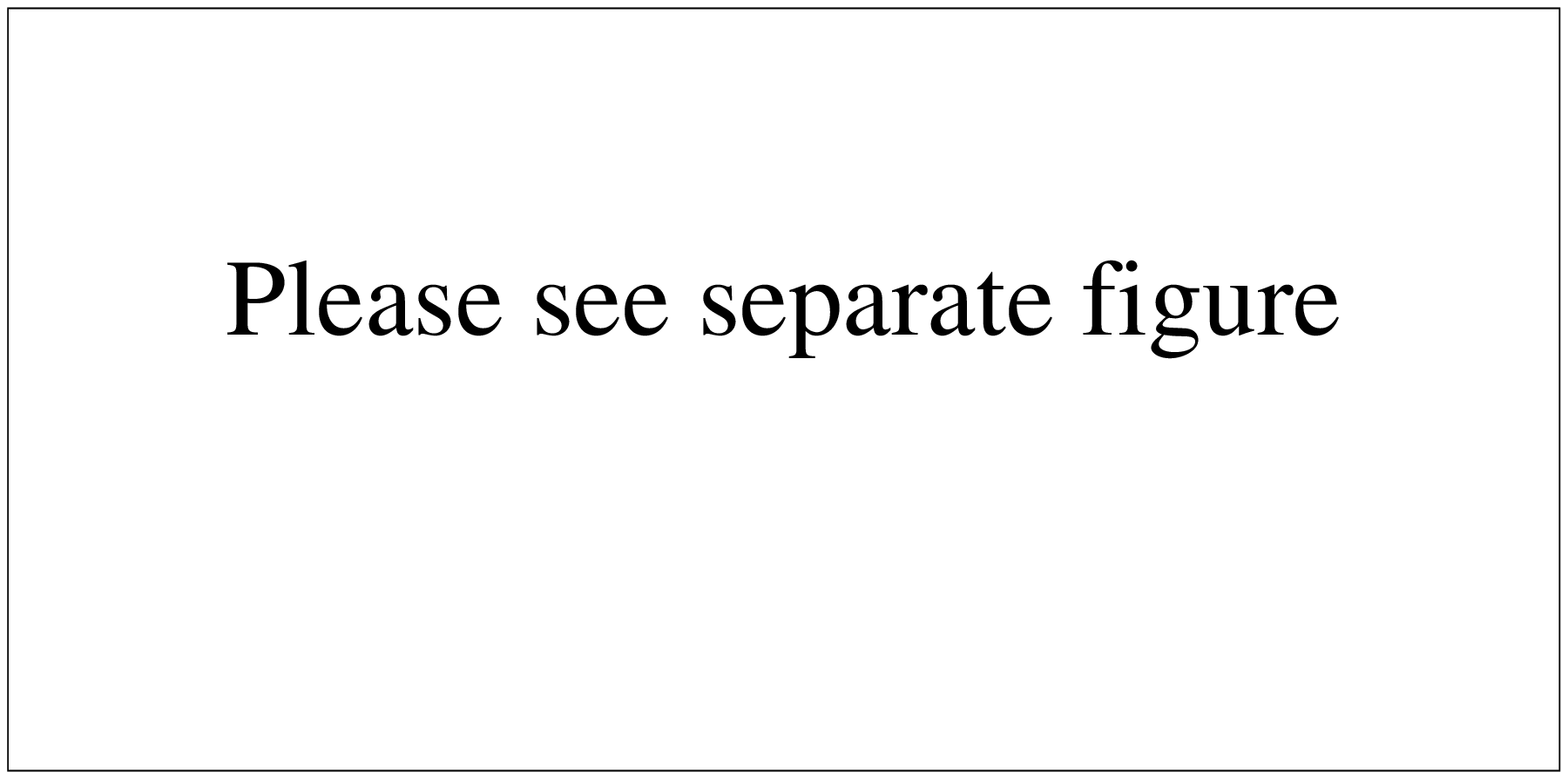,width=0.45\textwidth,angle=270}
\caption{\label{fig:3c323.1diag}
Outline of the nebula associated with 3C323.1, with diamonds marking
the 20 fibres used
to calculate average values of line intensity ratios given in
Table~\ref{tab:linerats}. The asterisk marks the
peak fibre, and the straight lines show the position angles of the
three cuts at 90\degmark, 120\degmark\ and 150\degmark\ along which
values of both intensity and radial velocity of the \oth line are
shown in Fig~\ref{fig:3c323.1rvint}. }
\end{figure}

\begin{figure}
% allrats.out
\psfig{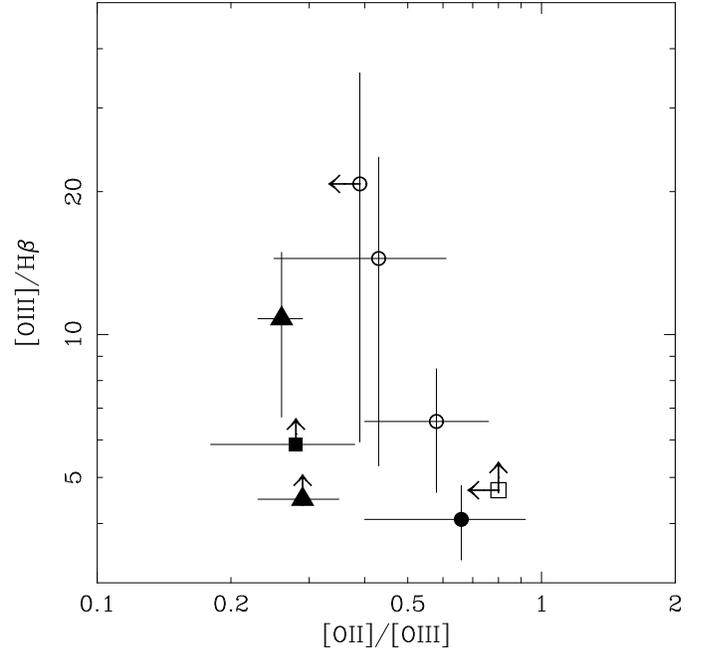}
\caption{\label{fig:allrats}
Line intensity ratios of [OIII]/\Hb plotted against those of
[OII]/[OIII] for the quasar nebulosities as given in Table~\ref{tab:linerats}. 
The ratios in the extended line emission of 3C323.1 are marked by a
solid circle, 4C11.72 by open circles, 
the limits to 3C215 by an open square, 
3C334 by solid triangles and 
3C281 by a solid square.}
\end{figure}

Redshifted \otw line emission is barely detected in the EELR, even when the
spectra are binned into larger 7-fibre cells, as it falls in the less
sensitive region of the combined chip$+$grism response. Narrow \Hb is also
too faint to be significantly detected outside of the nuclear regions in
individual fibres. We obtain an off-nuclear detection of each of [OII] and
\Hb only by summing the spectra from  20 fibres forming the
most extended region at the SE of the nebula (the fibres marked by
diamonds in Fig~\ref{fig:3c323.1diag}). Fitting the total spectrum
from this region enables us to obtain average line fluxes (although
[OII] had to be fixed at the same velocity width as the [OIII] and \Hb
complex) and thus intensity ratios of [OIII]/\Hb and [OII]/[OIII]
(Table~\ref{tab:linerats} and Fig~\ref{fig:allrats}). We calculate the
fractions of total line emission emitted from the nucleus (defined as
emitted by the central 19 fibres) in the both [OIII] and [OII] and
find them to be 69 and 52 per cent respectively
(Table~\ref{tab:nebprops}). These may well
represent underestimates of the actual fractions, as we do not take
into account any EELR that may be contributing emission along the line
of sight to the quasar nucleus.

\begin{figure}
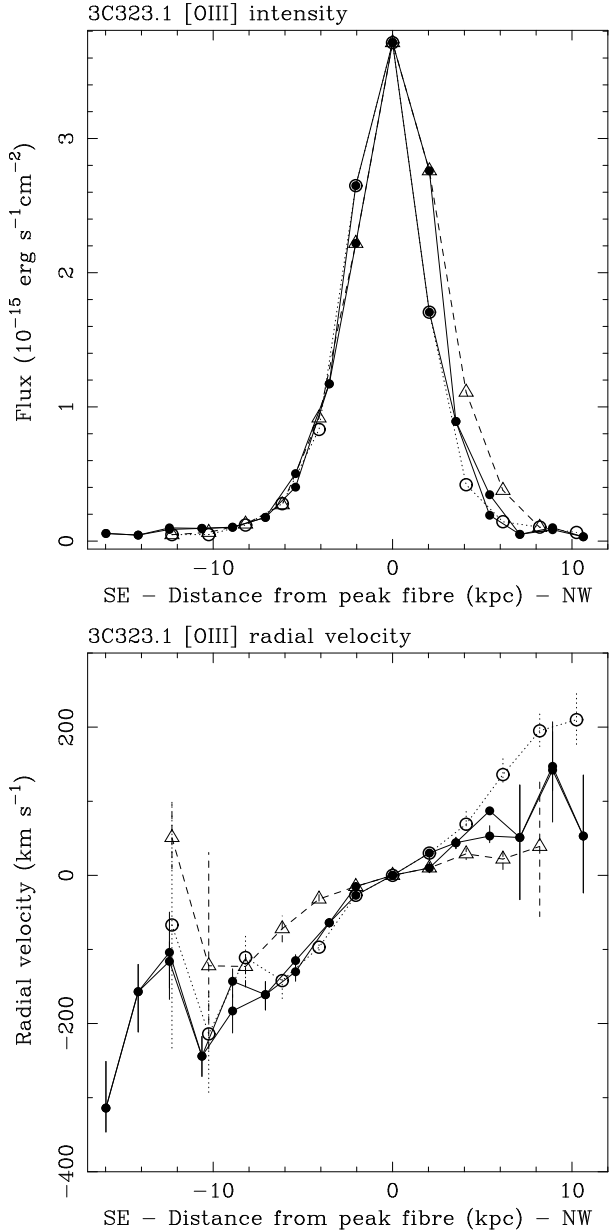

\vbox{
% /data/csc/argus/jundata/3c323.1/results/int.qdp
\psfig{figure=fig3t.ps,width=0.45\textwidth,angle=270} 
\vspace{0.2cm}
% /data/csc/argus/jundata/3c323.1/results/radvely.qdp
\psfig{figure=fig3b.ps,width=0.45\textwidth,angle=270}}
\caption{\label{fig:3c323.1rvint}
The intensity (top) and the radial velocity (below) of the [OIII] line emission
relative to the peak fibre along three cuts through the nebula in 3C323.1.
The dashed line marked by triangles is along the cut at 90\degmark, the
dotted line marked by open circles is that along 150\degmark, and the solid
lines and circles show the values along 60\degmark\ (note that some of the
values are double along the latter cut because it does not pass through only
a single row of fibres). The width of the PSF from the nuclear line intensity
varies according to whether the cut passes more along or across the rows in
the hexagon. The errors on the radial velocity map are the $\Delta\chi^2$=1
confidence values from the fits to the emission lines and do not take
systematics that may be introduced by the data reduction into account.
}\end{figure}

Fig~\ref{fig:3c323.1im} also shows the radial velocity of the \oth line
relative to that the peak fibre. The nebula appears to show a fairly regular
dipolar motion with amplitude of $\pm$200\kmps, albeit about an axis misaligned from that of the radio
source by approximately 25 degrees. The radial velocity values along cuts at
position angles of 90\degmark, 120\degmark and 150\degmark (as marked in
Fig~\ref{fig:3c323.1diag}) are shown in Fig~\ref{fig:3c323.1rvint}. The
linewidth (FWHM as measured from the fit to the [OIII] lines) is highest
around the nucleus, dropping to 200--400 \kmps in the extended emission to
the SE.

\begin{table*}
\caption{\label{tab:linerats} Average line ratios in the EELR of the quasars {\em except} for 4C37.43
whose line ratios are shown in Fig~\ref{fig:4c37.43bpt}.}
\begin{tabular}{rlcc}
        &                       &               & \\
Quasar  & Distance from nucleus & [OII]/[OIII]  & [OIII]/H$\beta$ \\
        & (kpc)                 &               &                 \\
\hline
3C323.1 & 14$\pm$4\  (SE)        & 0.7$\pm$0.3 & 4.1$\pm$0.7 \\
%         14.3 +3.9 -3.5 at proper distances!
% 0.66$\pm$0.26 & 4.08$\pm$0.73 \\
4C11.72   & 20$\pm$6  (E) & 0.6$\pm$0.2 & 6.6$\pm$1.9 \\ 
%! range from radius: 1406391.12 4894172.5
% 0.58 0.18  6.56 1.91
          & 24$\pm$7 (SE) & $<$0.4      & 20.8$\pm$14.8 \\
%          & 24$\pm$7 (SE) & $<$0.39       & 20.76$\pm$14.82 \\
%! range from rasius : 610105.438 2296924.75
                    & 22$\pm$5  (S) & 0.4$\pm$0.2 & 14.5$\pm$9.2\\ 
%                    & 22$\pm$5  (S) & 0.43$\pm$0.18 & 14.46$\pm$9.17\\ 
%! range from radius : 903748.375 2378899. 
% proper distances     20.1$\pm$5.5  E 
%                      24.1$\pm$7.3 SE 
%                      22.4$\pm$4.7  S 
3C215  & 13$\pm$5 (N)  &  $<0.8$ & $>4.7$ \\
%3C215  & 13.1$\pm$5.2 (N)  &  $<0.8$ & $>4.7$ \\
3C334        & 22$\pm$1 (outer)   & 0.3$\pm$0.1 & 10.8$\pm4.1$  \\
%3C334        & 22$\pm$1 (outer)   & 0.26$\pm$0.03 & 10.8$\pm4.1$  \\
%! from radius 2375410.25 3152453.5
%3C334         & 14$\pm$6 (inner)  & 0.29$\pm$0.06 & $>$4.5  \\
             & 14$\pm$6 (inner)  & 0.3$\pm$0.1 & $>$4.5  \\
%! from radius 17205756. 6125729.
%out 22.02 +1.52 -1.26 ie 1.39 
%in  13.96 -3.27 +2.55 ie 5.82
3C281  & 15$\pm$4 (NW)  & 0.3$\pm$0.1 & $>$5.9 \\    
%3C281  & 15$\pm$4 (NW)  & 0.28$\pm$0.10 & $>$5.87 \\    
% 3C281  14.7 4.2 
       &                       &               & \\
\end{tabular}
~~~~~~~~~~~~\\
\noindent Notes: The errors on the intensity ratios given are not the
extrema, but are propagated from the $\Delta\chi^2$=1 errors on the
intensity in the line fits. \\
\end{table*}

\subsection {4C11.72, $z$=0.323}
4C11.72 is a relatively compact (diameter $\sim75$\kpc), linear radio
source associated with a low-redshift quasar located in a small
cluster of galaxies (Gunn 1971; Robinson \& Wampler 1972; Yee \& Green
1984; Block \& Stockton 1991; Ellingson \etal 1994). The quasar lies
in a relatively undisturbed host galaxy (Hutchings \& Neff 1992), but
has an [OIII] emission-line region extended along position angle
100\degmark, with an orientation and a spatial scale comparable to the
radio source (SM87; Hutchings \& Crampton 1990; Durret \etal 1994).
4C11.72 is the best case amongst low-redshift quasars for an
association between the EELR and radio emission, although not all the
ionized gas is related to the radio structure. The EELR is clumpily
distributed, with two bright concentrations $\sim4$ arcsec south-east
of the quasar nucleus, spatially coincident with a radio lobe.
Spectroscopy of the EELR shows the gas to have an irregular and
chaotic velocity structure throughout the envelope, with a range in
velocities of a few hundred \kmps (Hutchings \& Crampton 1990; Durret
\etal 1994).
%The quasar
%shows associated absorption lines of CIV, SiIV and Ly~$\alpha$
%(Ellingson \etal 1994).

\begin{figure}
% alternatively, the better colour version is in /colour/4c11.72col.ps
% made from /data/csc/argus/jundata/4c11.72/results : colnewfig.idlcom and newerfig.idlcom
\psfig{figure=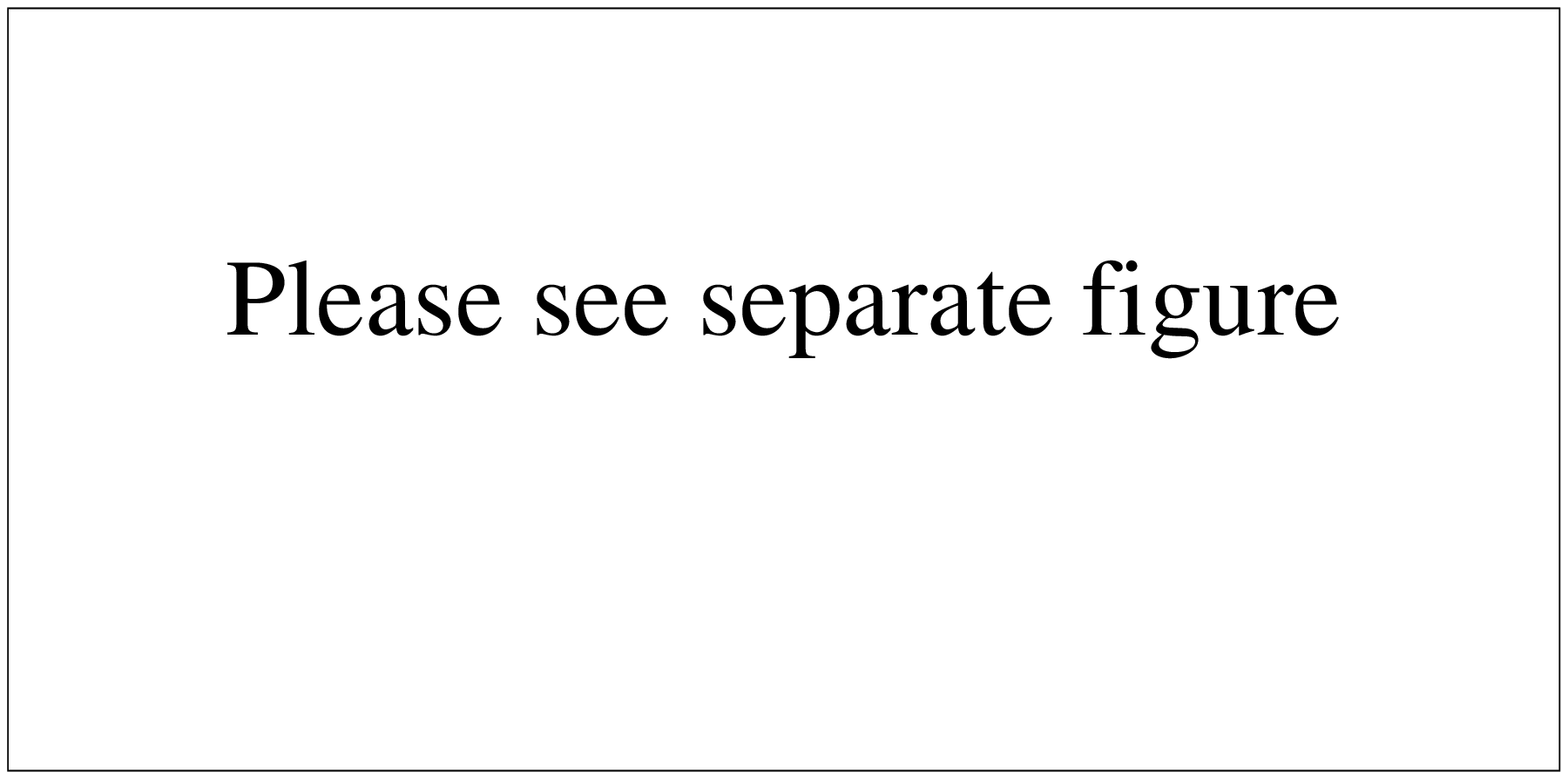,width=0.45\textwidth,angle=270}
\caption{ \label{fig:4c11.72im}
Reconstructed ARGUS images of 4C11.72 in the light of 
\oth line intensity (top left; on a scale of 0 (white) to a maximum (black) of
$2.09\times10^{-15}$\ergpspcmsq with minimum detection of 
0.02$\times10^{-15}$\ergpspcmsq); 
broad \Hb line intensity (lower left; 
scale 0--7.43$\times10^{-15}$\ergpspcmsq with minimum detection of 
0.10$\times10^{-15}$\ergpspcmsq); 
radial velocity relative to the nucleus over $-490$ to 360\kmps 
(top right) and FWHM (lower right; 0-1200\kmps) 
of the \oth emission line. 
The appropriate colour scale is shown to the right of each of the
kinematic maps. The position of the nuclear fibre is marked by a cross 
in these images, and the orientation of the radio source
axis is marked by a thick line in the radial velocity plot, 
 and those fibres where the line width was unresolved are
marked by a plus mark in the FWHM plot. 
The hexagonal ARGUS aperture is drawn around each image. 
North is to the top and east to the left, and 
2 arcsec corresponds to $\sim11.5$\kpc\  at the redshift of the quasar. }
\end{figure}

\begin{figure}
%\vspace{-1.0cm}
% /data/csc/argus/argus/jundata/4c11.72/results/diag.idlcom
\psfig{figure=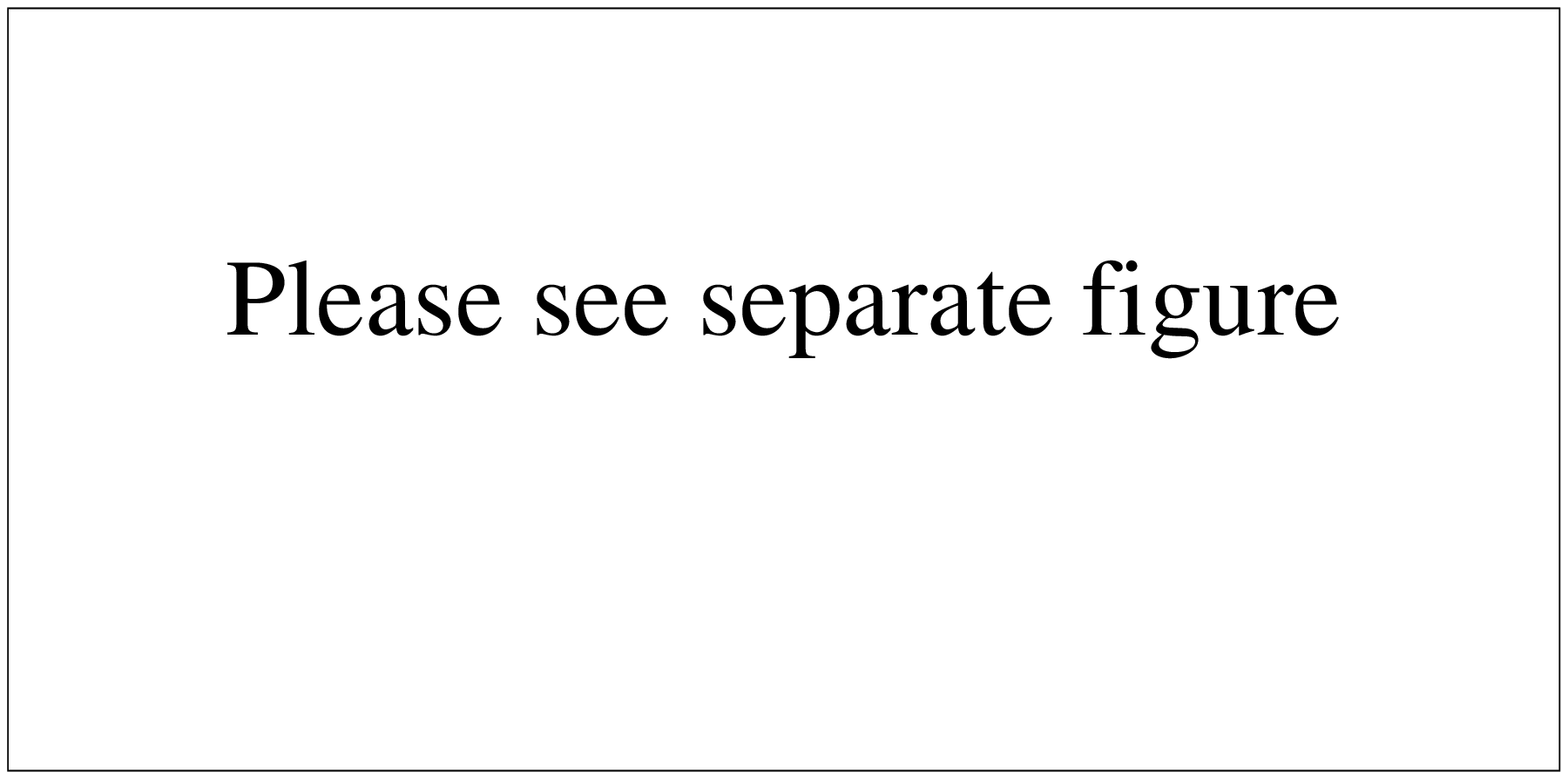,width=0.45\textwidth,angle=270}
\caption{\label{fig:4c11.72diag}
Outline of the nebula associated with 4C11.72, with the fibres used to
measure average values of line intensity ratios given in
Table~\ref{tab:linerats}. The fibres comprising the \lq East' 
region are marked by pluses, those in the 
\lq South-East' region are marked by diamonds, and in 
the \lq South' region are marked
by crosses. The asterisk marks the
peak fibre, and 
the straight lines mark the 
position angles of the cuts along 
150\degmark\ from which values of both
intensity and radial velocity are shown in Fig~\ref{fig:4c11.72rvint}. 
}
\end{figure}

The EELR extends right to (and probably beyond) the edges of the ARGUS
aperture to the south and north-west (Fig~\ref{fig:4c11.72im}). The
\oth line emission from the quasar is spatially extended along an
approximate south-east to north-west direction forming a rhomboidal
shape of approximate size 8$\times$6 arcsec ($\sim45\times$35\kpc; 1
arcsec corresponds to $5.7$\kpc\ at the redshift of the quasar). The
image of the nucleus of this quasar in both the light of
\oth and broad \Hb is asymmetric (this is a real variation in the
intensity of the emission lines between each fibre and not an artefact
caused by the differential response of fibres themselves). The quasar
is again at too low a redshift for the \otw to be detectable, as it
lies in the least sensitive part of the chip$+$grism response. We have
instead divided the south-east EELR into three (approximately) equal
wedges to the east, south-east and south (Fig~\ref{fig:4c11.72diag}),
well beyond the nuclear light spillover, and have summed the fibre
spectra to give an average spectrum in each of the three regions. The
\otw line was clearly detectable in the two of these regions, and we
obtain an upper limit to it in the SE region, fixing its width and redshift
to be the same as that fitted to the \oth line complex. The average line
intensity ratios derived are shown in Table~\ref{tab:linerats} and
Figure~\ref{fig:allrats}.

\begin{figure}
%~ ~ ~ ~ ~ \\
%\vspace{3cm}
%\hspace{1cm}
%\vspace{8cm}
% better (but double) colour version to be found in colour/4c11intandradcol.ps
%\psfig{figure=fig7.ps}
%\vspace{-2cm}
\psfig{figure=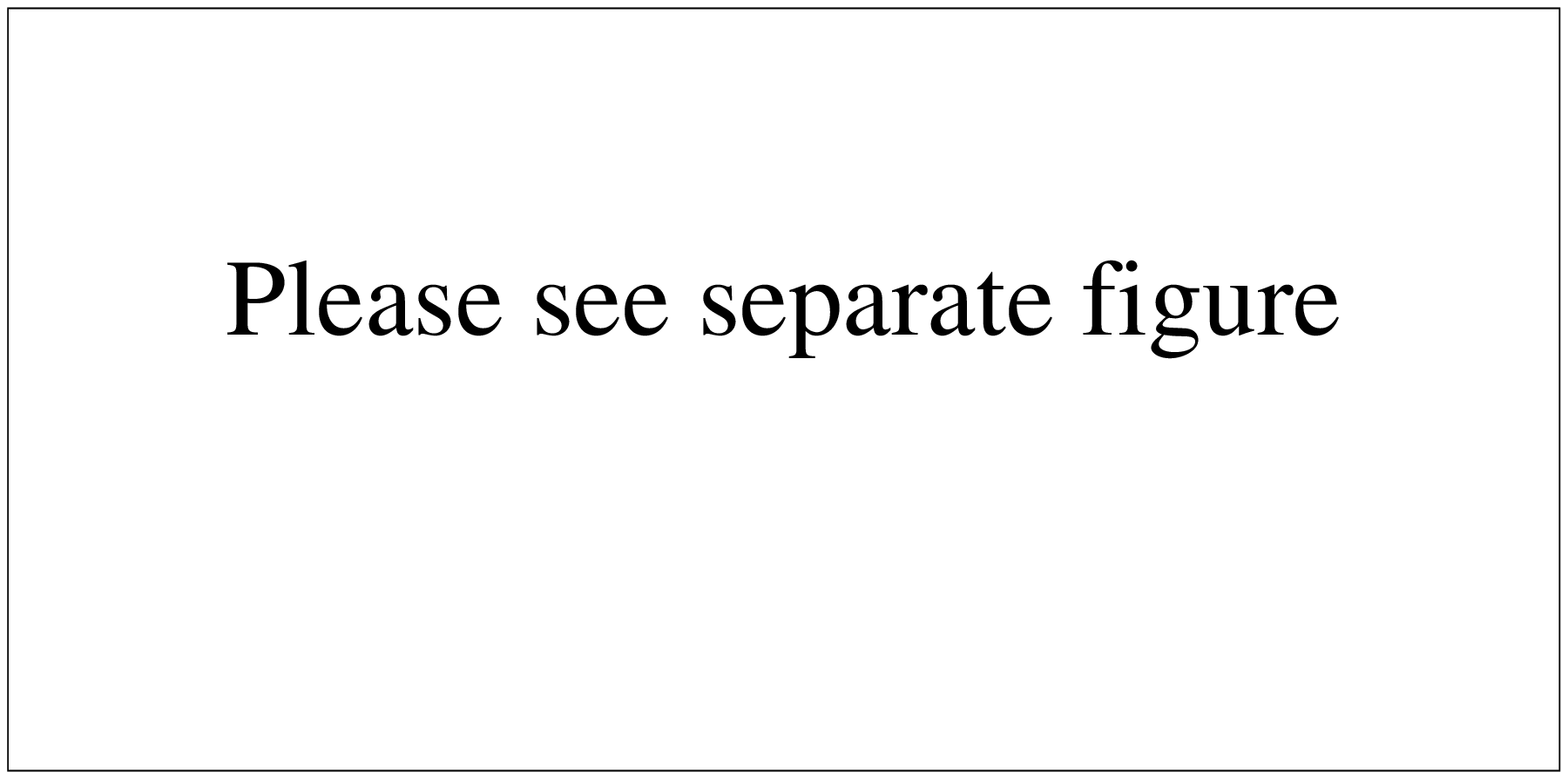,width=0.45\textwidth,angle=270}
\caption{\label{fig:intandrad}
The ARGUS image of the radial velocity structure of the
extended [OIII] line 
emission in 4C11.72. Contours of 4.86GHz radio emission from the 
radio source (from Miller \etal 1993) are superposed at the correct
orientation and scale. }
\end{figure}

At first sight, the radial velocity of the EELR appears to have an
approximately dipolar structure about an axis almost perpendicular to the
radio source axis. The [OIII] line is redshifted by about 200\kmps to
the north-west of the nucleus, but the south-eastern side is uneven,
with a region of blueshifted material with velocities ranging over
--200\kmps to --400\kmps to the south (Fig~\ref{fig:4c11.72im}). The
velocity structure we observe is in agreement with that found by
Hutchings \& Crampton (1990) and Durret \etal (1994). There is a
particularly close correspondence between the radio source structure
where it is co-spatial with the EELR to the south of the quasar
nucleus; the most strongly blueshifted gas is directly spatially
coincident with the double radio hotspot 2.4 arcsec south-east of the
quasar nucleus (Fig~\ref{fig:intandrad}). This correspondence naively
suggests a direct physical interaction between the radio and optical
plasmas in this region, but the linewidth of [OIII] remains relatively
narrow ($<500\kmps$) in this region, increasing to $\sim900$\kmps only
around the nucleus (Fig~\ref{fig:4c11.72im}). The ionization state of
the gas is, however, slightly higher in this region to the south than
that more to the east of the quasar nucleus (Fig~\ref{fig:allrats} and
Tab~\ref{tab:linerats}).

\begin{figure*}
\hbox{
% /data/csc/argus/jundata/4c11.72/results/o3int.plot
\psfig{figure=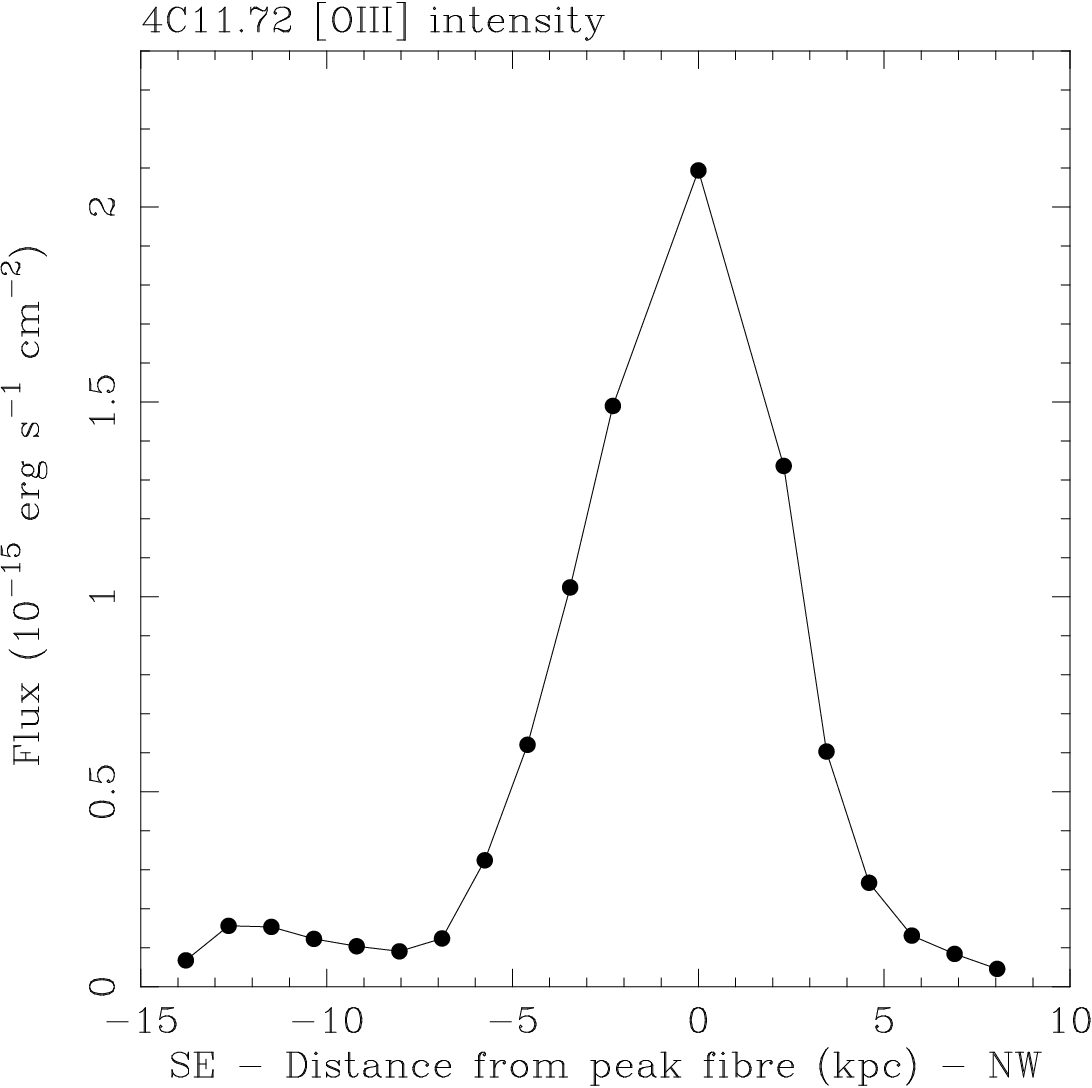,width=0.45\textwidth,angle=270} 
\hspace{0.2cm}
% /data/csc/argus/jundata/4c11.72/results/radvely.plot
\psfig{figure=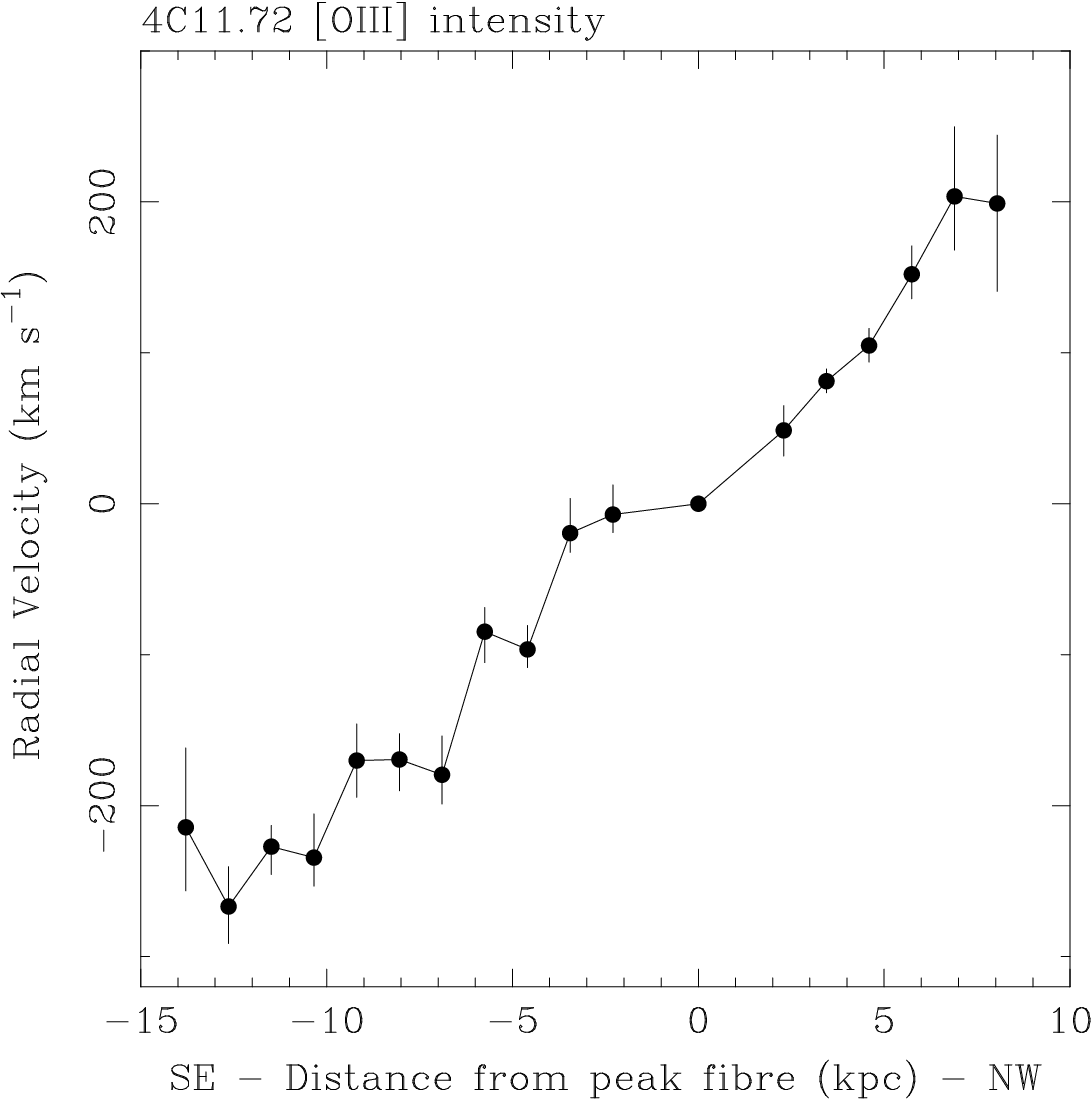,width=0.45\textwidth,angle=270}}
\caption{\label{fig:4c11.72rvint}
The intensity (left) and the radial velocity relative to the peak fibre (right) of the \oth line emission
along a cut following a position angle of
150\degmark\ through the nebula of 4C11.72. 
The errors on the radial velocity map are the $\Delta\chi^2$=1
confidence values from the fits to the emission lines and do not take
systematics that may be introduced by the data reduction into account.
}\end{figure*}

\subsection {4C37.43, $z$=0.371}

\begin{figure}
% colour version in /data/csc/argus/jundata/4c37.43/results/bestcolfig.ps
\psfig{figure=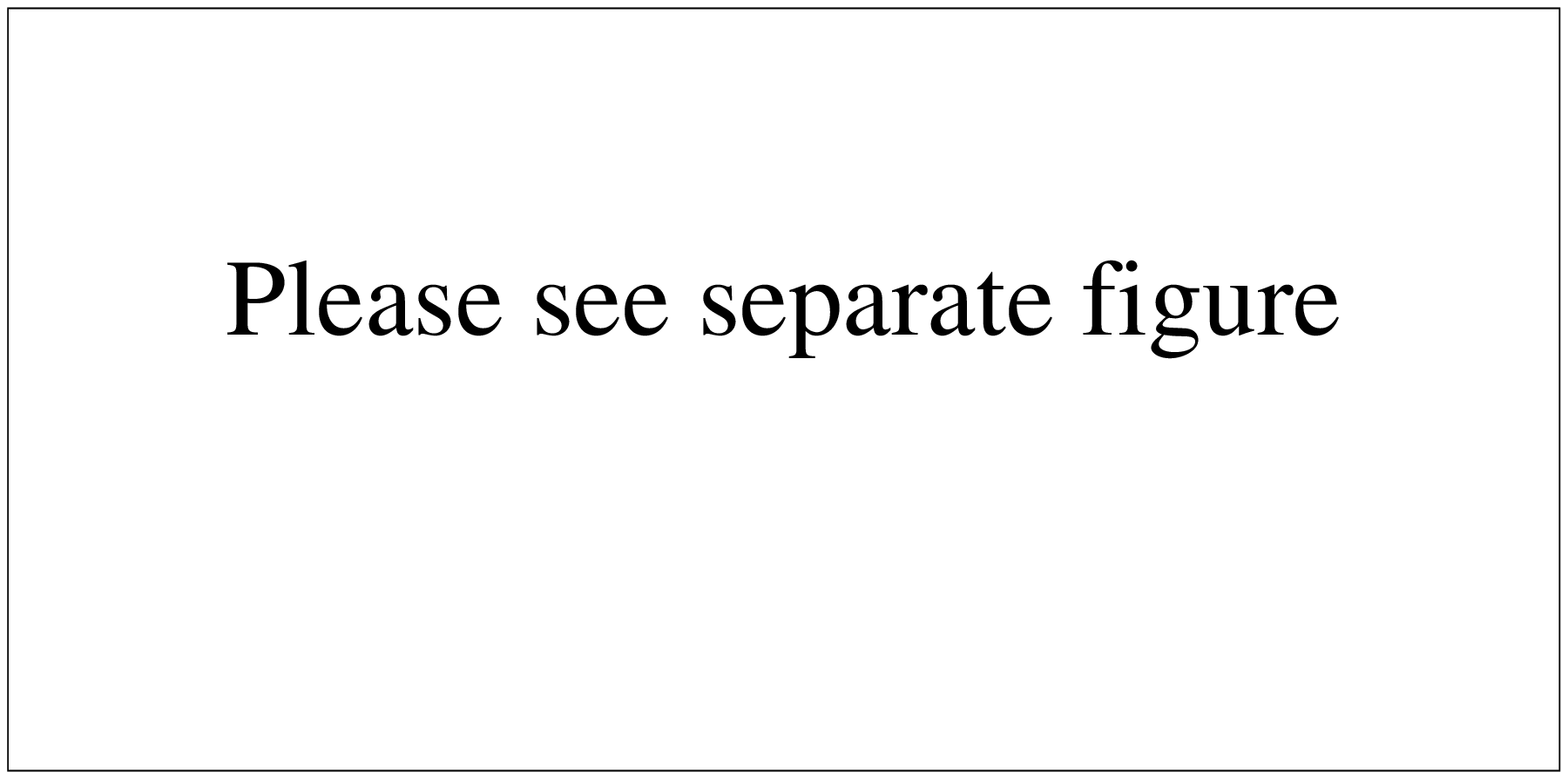,width=0.45\textwidth,angle=270}
\caption{\label{fig:4c37.43fig}
Reconstructed ARGUS intensity maps of 4C37.43 in the light of
\oth (plotted over the range 0.02(white)--1.3(black)$\times10^{-15}$\ergpspcmsq),
broad \Hb (range 0.29(white)--6.77(black)$\times10^{-15}$\ergpspcmsq),
and the ratio of [OIII]/\Hb (range 4--11). The radial velocity (from
-550 to 200\kmps with respect to the nucleus) and FWHM of the EELR
(0-750\kmps) as measured from the fits to the [OIII] line emission are
shown to the right, each with its appropriate colour bar. The outer
hexagon marks the ARGUS aperture for each image, and a cross marks the
nuclear fibre in the ratio and kinematic maps. The solid line through
the nucleus in the velocity map describes the direction of the radio
source axis, and plus marks on the FWHM map show those fibres where
the [OIII] line was unresolved. North is to the top of the image, east
to the left, and 2 arcsec corresponds to a diameter of 12.4\kpc\ at
the redshift of the quasar. }\end{figure}

4C37.43 has one of the most spectacular and luminous EELRs known 
around a low-redshift quasar, with a full extent of 200\kpc\ (SM87).
The EELR is distributed into two main condensates away from the quasar
itself; a large cloud centred 3.4 arcsec (21\kpc\ at a redshift of
0.371) directly east of the quasar nucleus, and a thin elongated cloud
running north-east to south-west centred 3.4 arcsec  to the north-west of
the quasar (Fig~\ref{fig:4c37.43fig}; SM87). Durret \etal (1994) show
that these two main regions have velocity fields 
which can each be ascribed to a simple rotating disk decoupled from
the rest of the line emission. 
The continuum structure
around the quasar is also complex, with a faint companion galaxy lying
10 arcsec east, perhaps linked to the quasar by a straight continuum \lq
bridge', and an elongated galaxy just beyond the emission cloud at
$\sim6$ arcsec to the north-west (Block \& Stockton 1991). The
distribution of the continuum- and line-emitting material appears
completely uncorrelated; in particular both clouds have no associated
continuum. The quasar may be associated with a small group of galaxies
(Yee \& Green 1984; Ellingson \etal 1994).
The radio source associated with the quasar is a straight
and symmetric FR~II double source, with the hotspots separated by
$\sim60$ arcsec (370\kpc) at a position angle of 110\degmark\ (Miller,
Rawlings \& Saunders  1993). 

\begin{figure}
\psfig{figure=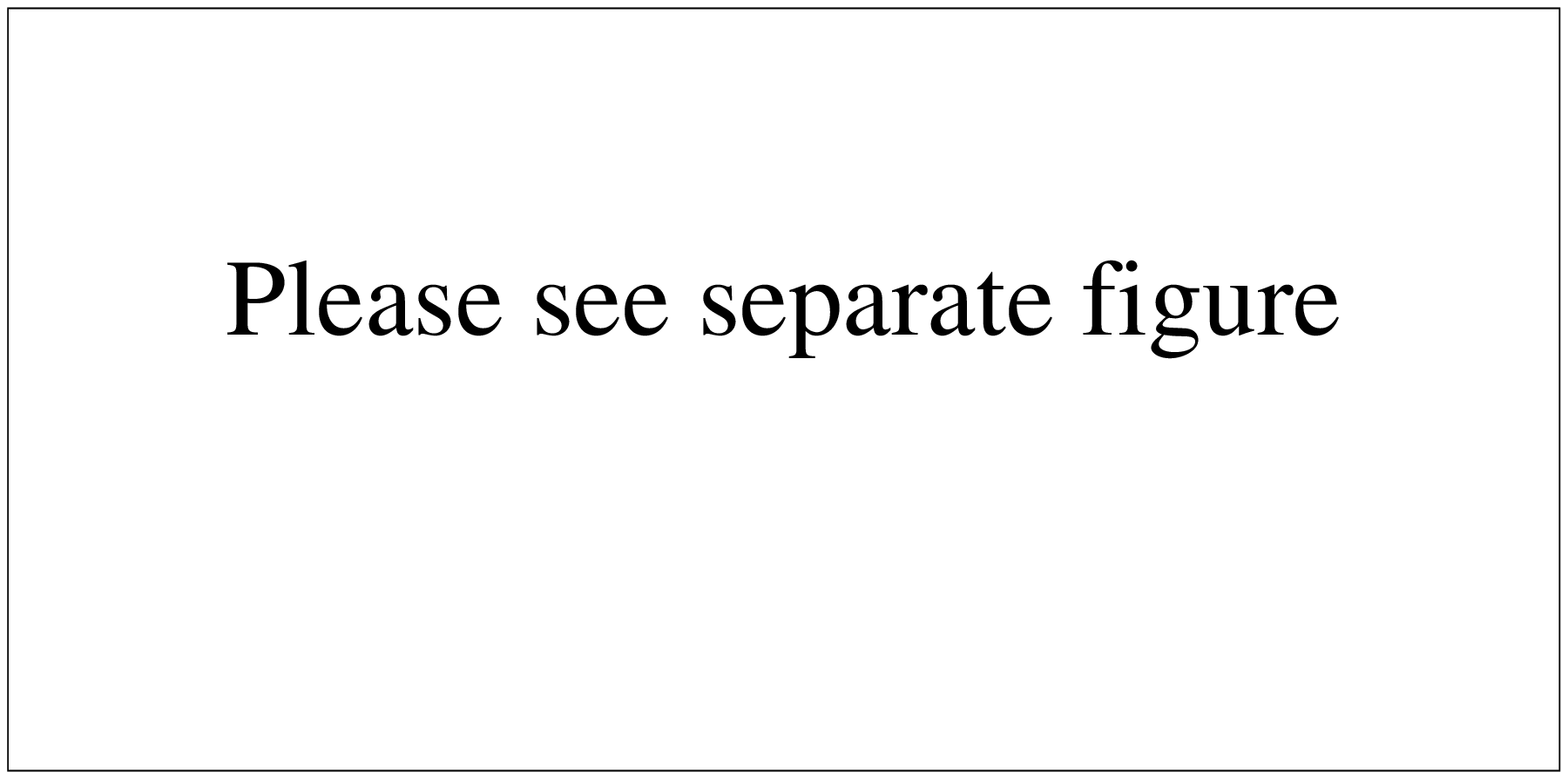,width=0.45\textwidth,angle=270}
\caption{ \label{fig:4c37.43row12}
The spectrum of each individual fibre in row 12 of the 4C37.43 
observation (the row crossing the
nucleus; Fig~\ref{fig:4c37.43fig}) 
over the wavelength range of 6825--6925\AA, covering the spectral region
of redshifted \oth; wavelength increases to the right in each panel. Fibres are numbered from east to west across
the aperture, where fibre 17 marks the nuclear peak. Fibres 6--11 cover the 
brightest region of the large eastern cloud, fibres 14--20 show the
nuclear region where broad \Hb is detected, and fibres 25-27 cover the
southern end of the north-western cloud. The solid vertical line in each plot 
marks the observed position of the \oth line at 6874\AA\ in the nuclear
fibre, to show the blueshift from that position off-nucleus. Successive
fibres are centred 0.4 arcsec apart. 
Fibres 15--21 are plotted on a flux scale of
0--1.3$\times10^{-16}$\ergpcmsqps, and all other fibres on a flux scale of 
0--2$\times10^{-17}$\ergpcmsqps.  }
\end{figure}

\begin{figure}
% /data/csc/argus/jundata/4c37.43/results/o3int.plot
\psfig{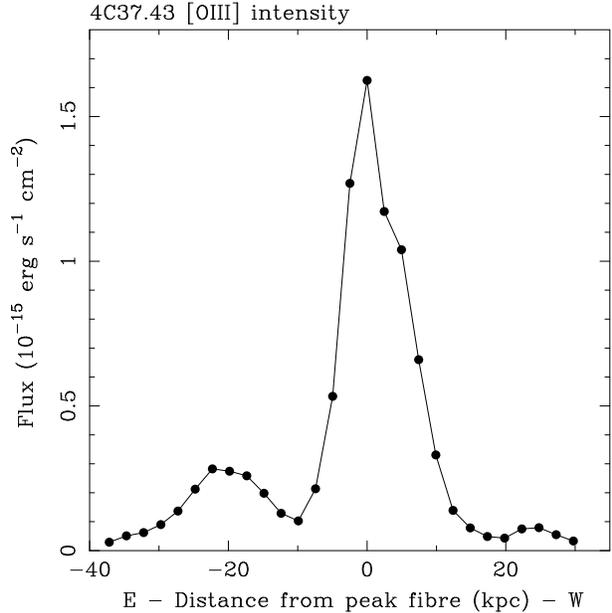} 
\caption{\label{fig:4c37.43int}
The intensity of the \oth line emission relative to the peak fibre
along an E-W cut through the nebula of 4C37.43. }\end{figure}

We centred our ARGUS aperture directly on the quasar, covering both
the eastern and north-western gas clouds, but not the extended
continuum structure around this quasar. Fig~\ref{fig:4c37.43fig} shows
our reconstructed ARGUS image of the quasar in the light of both \oth
and broad \Hb (to show the nuclear/instrumental PSF). The off-nuclear
condensates are apparent in [OIII] (see also
Figs~\ref{fig:4c37.43row12} and  \ref{fig:4c37.43int})
 and both they and the quasar seem
to be embedded in a low-level diffuse emission. 

\begin{figure*}
\hbox{
% /data/csc/argus/jundata/4c37.43/results/width.plot
\psfig{figure=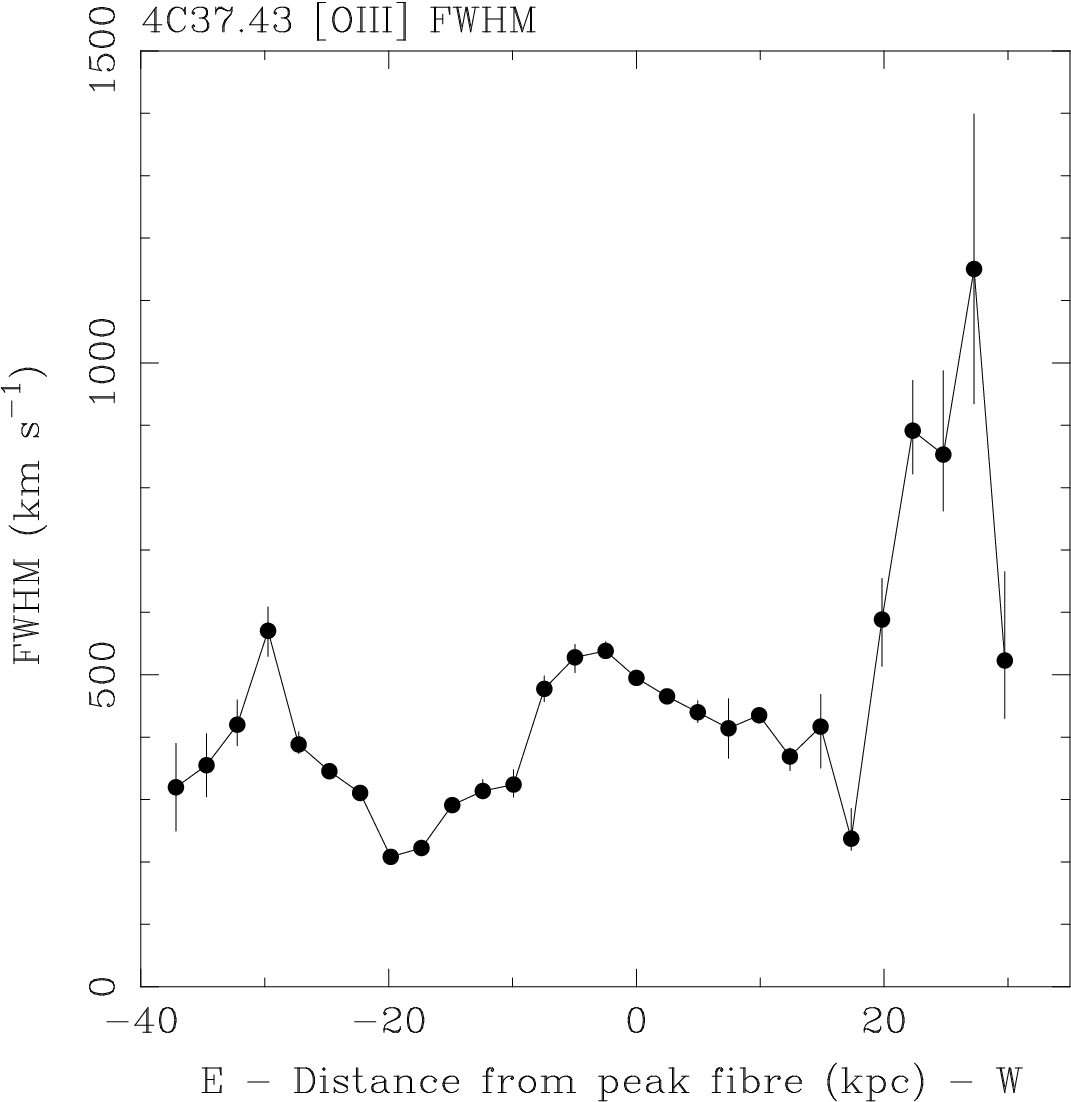,width=0.45\textwidth,angle=270}
\hspace{0.2cm}
% /data/csc/argus/jundata/4c37.43/results/radvely.plot
\psfig{figure=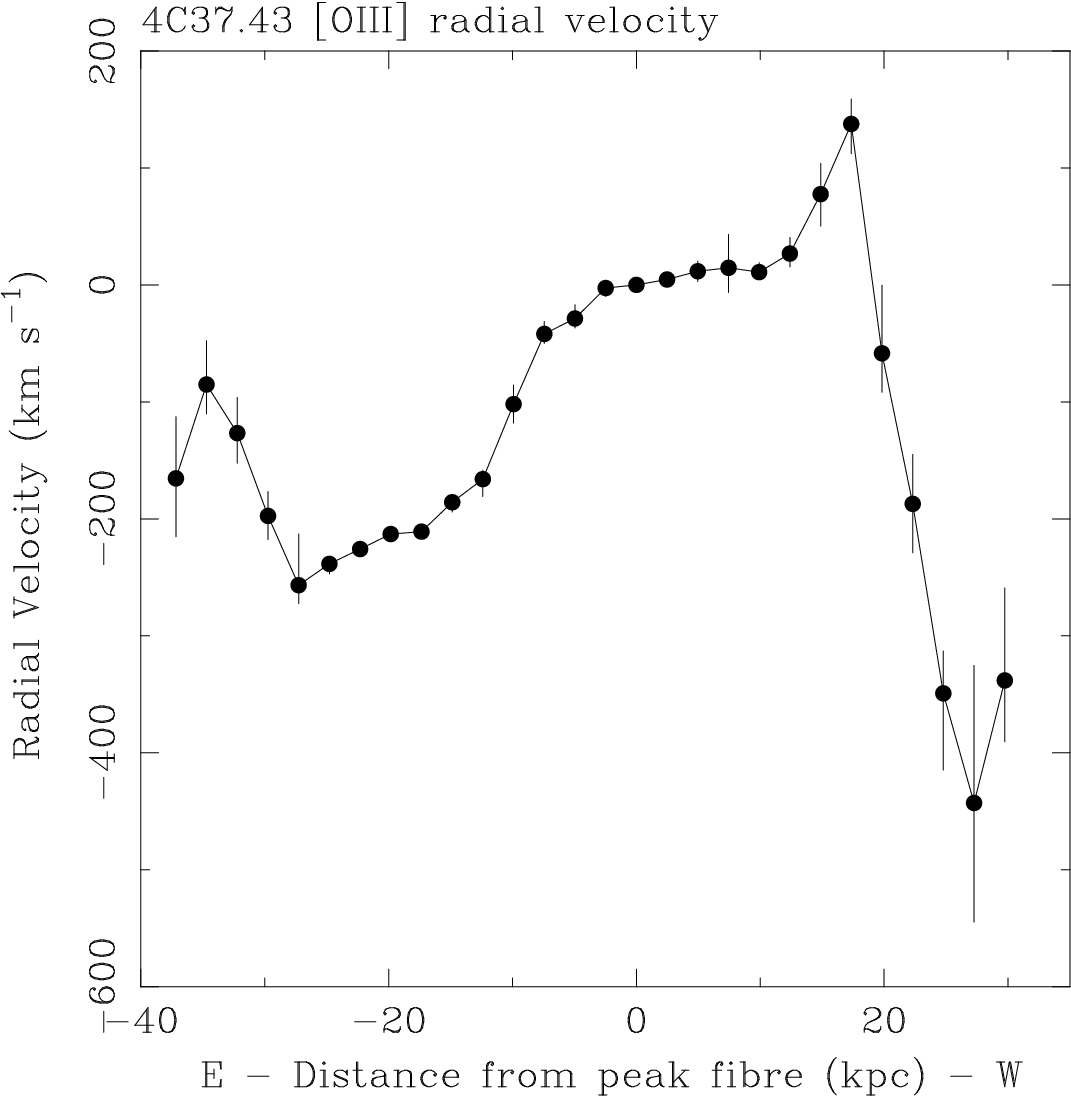,width=0.45\textwidth,angle=270}}
\caption{\label{fig:4c37.43rvfwhm}
The FWHM (left) and the radial velocity relative to the peak fibre (right) of
the [OIII] line emission along an E-W cut
through the nebula of 4C37.43. The errors on the radial velocity map
are the $\Delta\chi^2$=1 confidence values from the fits to the
emission lines and do not take systematics that may be introduced by
the data reduction into account. }\end{figure*}

The radial velocity map of the [OIII] line emission
(Fig~\ref{fig:4c37.43fig}) is again very consistent with that
published from field spectroscopy by Durret \etal (1994). The eastern
cloud is blueshifted by $\sim$200\kmps relative to the nucleus, and
the north-western cloud by $\sim400$\kmps
(Figs~\ref{fig:4c37.43row12}, \ref{fig:4c37.43rvfwhm}), although both
show a slight gradient across the cloud, increasing to $-$300 and
$-$500\kmps respectively towards the south. 

It is interesting to
compare the location and velocity gradient of the EELR to the radio
source structure, as both major condensations cross the radio axis. In
particular, the most blueshifted regions in each cloud spatially
coincide with the radio source axis. The nature of any
relation between the radio source and the optical line emission is far
from obvious, as the radio lobes lie far beyond the visible
emission-line region. In addition, the FWHM of the [OIII] line is
lower in the eastern cloud than at the nucleus ($<300$\kmps as opposed
to 500-600\kmps on-nucleus; Fig~\ref{fig:4c37.43rvfwhm}), suggesting
little turbulence of the gas here.  The off-nuclear gas is at much
higher FWHM in the southern end of the elongated north-western gas
cloud, however, reaching widths of 700--800\kmps where it crosses the
 radio source axis (Figs~\ref{fig:4c37.43fig}, \ref{fig:4c37.43rvfwhm}). This
is also the region in the EELR showing the largest blueshift
(--500\kmps). 

\begin{figure}
% fig is in /data/csc/argus/jundata/4c37.43/newcloudy/newbpt.out
\psfig{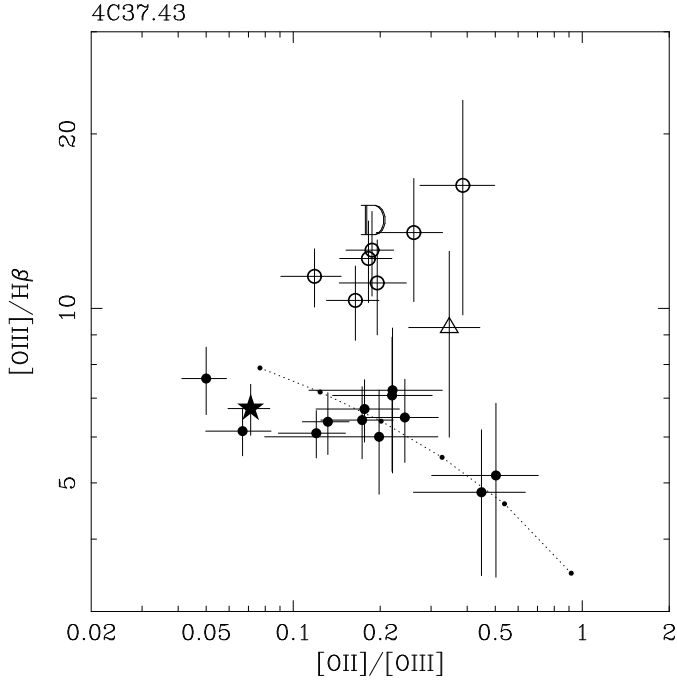}
\caption{\label{fig:4c37.43bpt}
Spatially resolved line intensity ratios for the large binned cells
(each of 7 fibres) for 4C37.43. The nuclear fibre is marked by a star
marker, the ratios at the position of the eastern cloud are marked as
open circles, and an average ratio at the north-western cloud is
marked as a large triangle. The remaining points (solid circles) are
those in the diffuse emission surrounding the nucleus. For comparison, the
average ratios measured for the eastern cloud by Durret \etal 1994 are
marked by \lq D'. The plotted line shows the range of predictions from
photo-ionization calculations in CLOUDY, for a fixed density of
100\pcmcu and an ionization parameter decreasing from -1.75 to -3 from left to
right, assuming a blue bump temperature of
$2.25\times10^5$\K. 
}\end{figure}

Narrow \Hb is detectable in individual fibres covering the core of the
eastern emission-line cloud, and the [OIII]/\Hb intensity ratio here
is markedly higher than around the nucleus (Fig~\ref{fig:4c37.43fig}).
Due to the combination of low quantum efficiency of the Loral~3
detector and the O300 grism at blue wavelengths, the
\otw emission line is too faint to detect in individual fibre spectra
covering the extended line emission. Instead we have summed the fibres
into larger cells each of seven fibres to gain a better detection of
[OII] off-nucleus. Where all of
\otwn, \oth and narrow \Hb emission lines are detected within our
aperture, we have formed the line ratios and plot them in a 
standard diagnostic diagram (Fig~\ref{fig:4c37.43bpt}). The observed
line ratios in the eastern cloud agree well with the average values
obtained by D94 and CF89. 

\subsection {3C215,   $z$=0.412}
3C215 lies in a densely clustered environment (Hintzen 1984;
Ellingson, Yee \& Green 1991) and the radio source has a very complex
structure. A knotted jet emerges from the nucleus to the east (pa
77\degmark), but the other main radio features on this side lie along
an angle of 100\degmark\ from the nucleus, as if the jet had been
deflected by about 40\degmark\ toward the south-east after 3.6 arcsec
(Bridle \etal 1994). The lobe of this and the other side of the radio
source are both very distorted. CF89 showed from long-slit
spectroscopy that the oxygen line emission from the quasar was
extended toward the north. Narrow-band imaging in the light of
redshifted [OII] shows an extension in the nebula for about
$\sim2$ arcsec along a position angle of 30\degmark\ (Hes \etal 1996).
Narrow-band [OIII] imaging by Hutchings (1992) did not show the host
nebulosity, but indicated the presence of three emission-line clouds
at further radii, associated with the radio source structure in some
way.

\begin{figure}
\psfig{figure=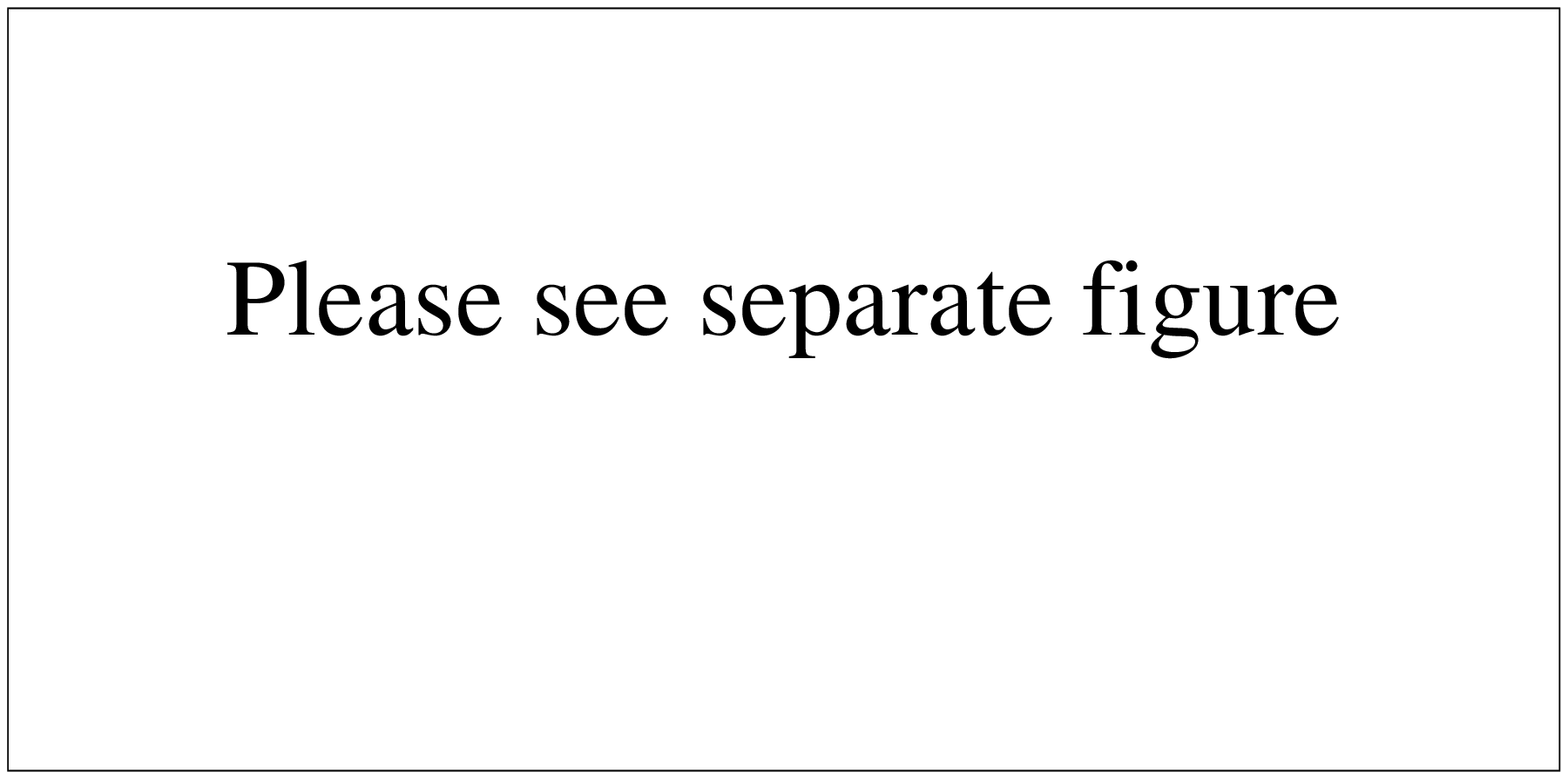,width=0.45\textwidth,angle=270}
%\psfig{figure=fig14.ps}
%\vspace{-5.0cm}
\caption{\label{fig:3c215fig}
Reconstructed ARGUS images of 3C215 in the light of 
\oth (left; range of 0-0.90$\times10^{-15}$\ergpspcmsq, where
0.01$\times10^{-15}$\ergpspcmsq is the minimum detection), 
 and in the 
red continuum  (right; range of 0.02--0.37$\times10^{-15}$\ergpspcmsq,
where 0.03$\times10^{-15}$\ergpspcmsq is the minimum detection). 
Note that in this quasar image is oriented through 45\degmark\ with
respect to the other quasars in this paper, so north-west is to the top
of the plot, and north-east to the left. 2 arcsec corresponds to a 
distance of 13\kpc\ at the redshift of the quasar. The solid line in the
[OIII] image indicates the direction of the jetted side of the radio
source and the location of its bend with the same scale and
orientation as the image (see text for details).  }
\end{figure}

The spectrum of the broad components to the Balmer lines within the
nucleus of this quasar are particularly difficult to fit in QDP, as
there appears to be a large blueshifted component to the broad lines
(Brotherton 1996). We have fitted the [OIII] and narrow \Hb complex,
without using a complicated fit to the broad \Hb line. Thus instead of
quantifying the nuclear PSF for this observation by means of the broad
\Hb line intensity, we have used a region of nearby featureless
continuum between 7200--7500\AA\ (observed; corresponding to
5100--5312\AA\ at rest wavelengths). In the light of \oth, this quasar
shows only a marginal extension out to a radius of 2.5 arcsec (17\kpc) directly 
towards north
(Fig~\ref{fig:3c215fig}; note that the image of this quasar is rotated
by 45\degmark\ compared to the rest of the quasars in this paper), but
it is not extended in the light of either \otw or narrow
\Hbn. There is no substantial velocity shift (ie $<150$\kmps with respect
to the nuclear velocity) anywhere within the [OIII] nebula. 
We obtain only
limits to the [OII] and narrow \Hb line emission from the sum of
eleven fibre spectra comprising this northern extension, and thus only
limits to the intensity ratios (Table~\ref{tab:linerats}). These
ratios are, however, consistent with the previous findings of CF89 who
measured [OII]/[OIII] of 0.5$^{+0.4}_{-0.2}$ at 29$\pm6$\kpc\ and
1.3$^{+4.0}_{-0.6}$ at 40\kpc\ from a long slit spectrum oriented in
the same direction from the nucleus. There is no obvious structure in
the FWHM of the \oth line across the system. 

\subsection {3C334,   $z$=0.555}
3C334 lies in a clustered environment (Hintzen 1984; Yee \& Green
1987), and is associated with a large double radio source with a very
strong jet to the south-east of the nucleus. CF89 showed from
long-slit spectroscopy that the oxygen emission lines are extended
north-south of the quasar. Hes \etal (1996) claim that the light of
redshifted [OII] is closely aligned with the direction of the strong
radio jet (see also Hes 1995). Imaging in [OIII] by Lawrence (1990),
however, shows the strongest line emission extended out to the north
of the quasar by several arcseconds.

\begin{figure}
\psfig{figure=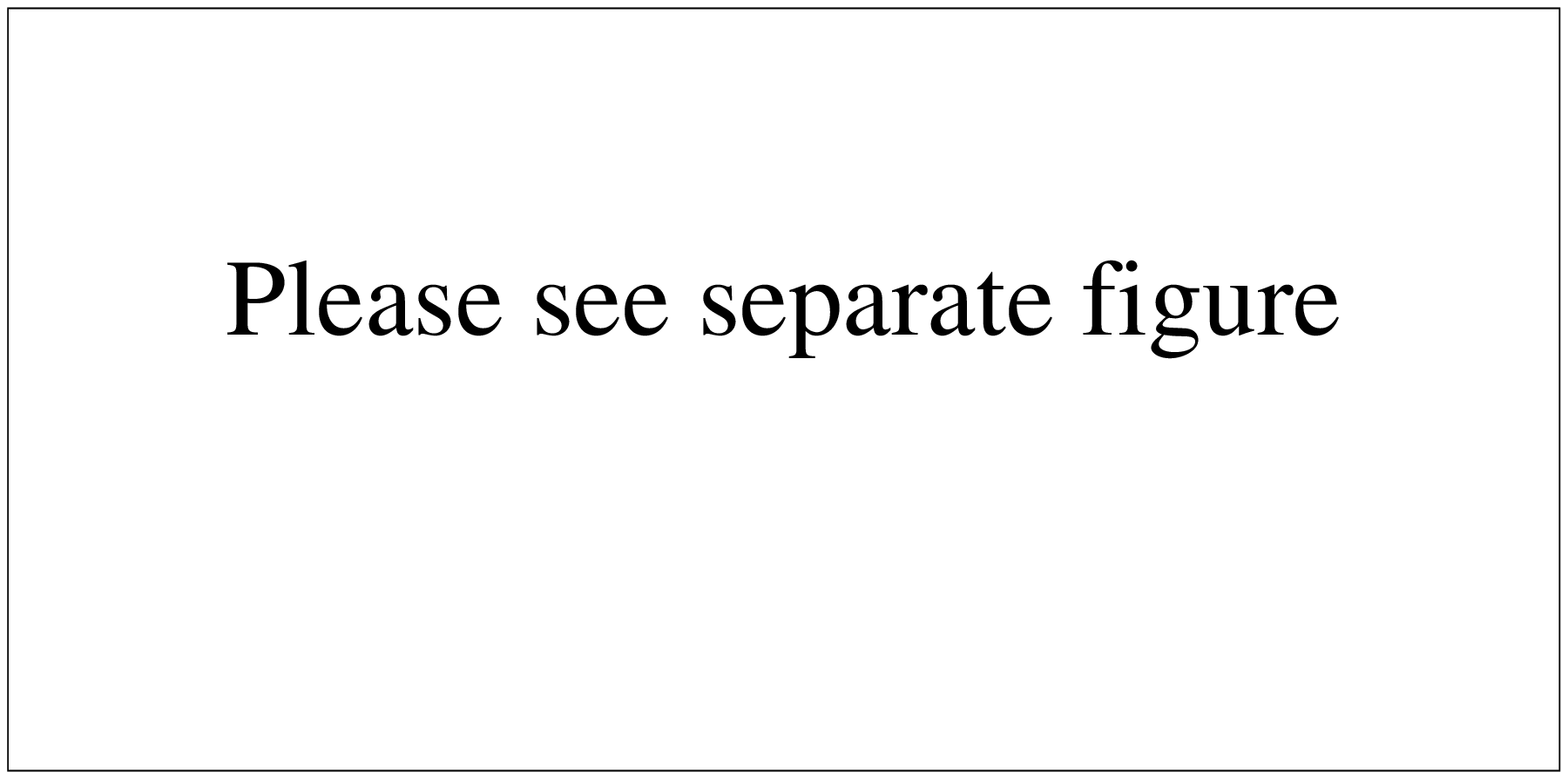,width=0.45\textwidth,angle=270}
%\psfig{figure=fig15.ps}
% better colour version in colour/3c334col.ps
\caption{\label{fig:3c334fig}
ARGUS images of 
3C334 in the light of \oth (range shown of
0--2.66$\times10^{-15}$\ergpspcmsq, where the minimum detection is 
0.03$\times10^{-15}$\ergpspcmsq); 
\otw (range shown of $0-0.40$\ergpspcmsq, where the minimum detection is
0.02$\times10^{-15}$\ergpspcmsq);
narrow \Hb (range shown of $0-0.31$\ergpspcmsq, where the minimum detection is
0.01$\times10^{-15}$\ergpspcmsq);
broad  H$\beta$ (range shown of $0-6.33$\ergpspcmsq, where the minimum
detection is
0.29$\times10^{-15}$\ergpspcmsq); and 
in radial velocity (over the range of $^{+150}_{100}$\kmps\ relative to the
nucleus). 
North is to the top and east to the
left, and the scale bar of 2 arcsec represents
a distance of $\sim$15\kpc\ at the redshift of the quasar. The solid line in the 
velocity map shows the direction of the jetted side of the radio source,
and the dashed line the direction of the radio counter-jet. 
}\end{figure}

Fitting the [OIII]$+$H$\beta$ complex in 3C334 requires special care,
as two atmospheric absorption bands around 7600\AA\ cut into the red
wing of the broad H$\beta$ emission component, between the narrow
H$\beta$ and [OIII]$\lambda$4959 emission lines. The absorption bands
were included in the fit as two gaussians when visible against the
broad H$\beta$ and continuum. The narrow emission-line gas
(principally [OIII]) is clearly extended into a plume stretching
4 arcsec ($\sim30$\kpc) to the north of the quasar
(Fig~\ref{fig:3c334fig}), with little sign of the extension to the
south-east (and out of the aperture).  It is less extended in \otw and
narrow \Hb. The northern plume is slightly redshifted (by about
100\kmps) and at a lower linewidth (around 350$\kmps$) than the gas
closer to the quasar nucleus, with the whole nebula forming
essentially a static system. The nuclear line emission has high
linewidths (800-1000\kmps), and is very slightly blueshifted to the
east of the nucleus.

The images in [OII] and [OIII] peak in neighbouring fibres due to slight
atmospheric refraction during the observation. Thus we have been careful
in forming line intensity ratios in the northern extended emission,
forming average ratios from two large cells within the nebula where the
extended emission in both lines is clearly significant. We average seven
fibres centred at a projected distance of 14\kpc\ north of the nucleus
(the \lq inner' region) 
and another 3 fibres 22\kpc\ north (the \lq outer' region: Table~\ref{tab:linerats}). 

\subsection {3C281,   $z$=0.599}
3C281 is associated with a large (330\kpc\ lobe separation) linear
steep spectrum radio source (Bogers \etal 1994), whose host quasar is
known to lie in a rich cluster (Yee \& Green 1987). Long-slit
spectroscopy by Bremer
\etal (1992) showed it to have extended [OII] and [OIII] emission out
to 5--8 arcsec to the south of the quasar.

\begin{figure}
\psfig{figure=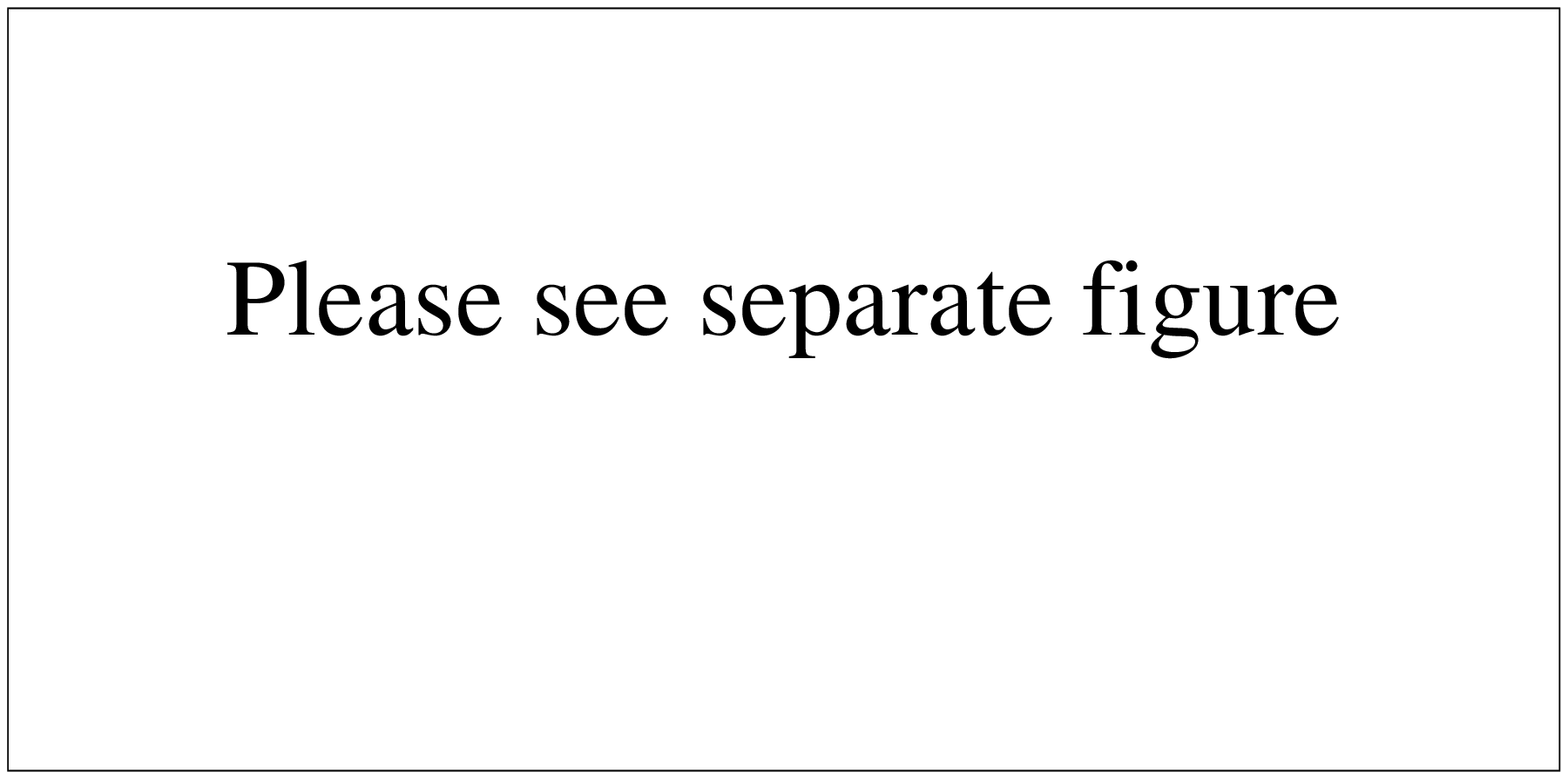,width=0.45\textwidth,angle=270}
\caption{\label{fig:3c281fig} 
ARGUS images of 3C281 in the light of
\oth (on a scale of 0-0.69$\times10^{-15}$\ergpcmsqps
with a minimum  value of 0.02$\times10^{-15}$\ergpcmsqps) 
\otw (on a scale of 0-0.15$\times10^{-15}$\ergpcmsqps
with a minimum value of 0.02$\times10^{-15}$\ergpcmsqps)
and broad \Hb (0-2.24$\times10^{-15}$\ergpcmsqps
with a minimum value of 0.09$\times10^{-15}$\ergpcmsqps). 
North is to the top and east to the left in these images, and the solid
line in the [OIII] image describes the direction of the radio source axis. 
2  arcsec   corresponds to a distance of 15\kpc\ at the redshift of the
quasar.}
\end{figure}

Our image shows the [OII] and [OIII] line-emitting gas only marginally
extended, out to 2.5 arcsec NW (corresponding to a projected distance of
19\kpc) from the quasar nucleus (Fig~\ref{fig:3c281fig}). Curiously we do
not find any emission to the south of the quasar, in contradiction to the
spectroscopic findings of Bremer \etal (1992). The orientation of the slit
in Bremer \etal is not in doubt (given another two objects also aligned
within the slit), but further direct comparison to our results is hampered
by the lack of a detection limit in that paper.
%******* {\em cant really figure out why, Bremer only shows relative flux
%and seems to have got orientation about right. Am confused} ******
The gas in the North-west extended region is redshifted by about 300\kmps relative to the
nucleus, and there is no discernible structure to the linewidth 
across the EELR. We sum seven fibres in this off-nuclear region to measure
average line intensity ratios for the EELR (Table~\ref{tab:linerats}).

\begin{table*}
\caption{Properties of the line emission from the quasars  and the radio
source. \label{tab:nebprops}}
\begin{tabular}{lllllllllc}
& & & &  & & & & & \\
Quasar &  L([OII]) & L([OIII])  & ${{\rm L}_{\rm n}}\over{{\rm L}_{\rm tot}}$([OIII]) 
& ${{\rm L}_{\rm n}}\over{{\rm L}_{\rm tot}}$([OII])  & p.a. & max
diameter & p.a.  & max diameter & Ref  \\
       & ($10^{43}$\ergps) & ($10^{43}$\ergps) & (\%)                            
& (\%)  & [OIII] (\degmark)  & [OIII] (kpc) & radio (\degmark) & radio (kpc) & \\
\hline
%Q      & L(o2) & L(o3) & Ln/Ltot o3  & Ln/Ltot (o2) & pa (o) & max diam (o) & pa (r)      & max diam (r) \\
3C215   & 0.05  & 0.54  & 85          & 90           & 10     & 25 &         77$^{\ast,\dag}$ (328) & 205 & Br \\ 
3C281   & 0.25  & 1.86  & 64          & 73           & 290    & 29
& 11     & 330 & Bo    \\
4C37.43 & 0.11  & 3.10  & 33          & 61           & 105    &
62$^{\ddag}$ & 110    & 375 & Mi    \\
3C323.1 & 0.13  & 1.63  & 69          & 52           & 115     & 32           & 20     & 360 & Bo  \\% Bogers etal at 1.4GHz 
3C334   & 0.62  & 5.13  & 68          & 63           & 355    & 41
& 140$^{\dag}$ (310)& 460 & Br \\
4C11.72 & 0.55  & 1.89  & 37          & 29           & 110    &
54$^{\ddag}$ & 138    & 75    & Mi    \\
& & & &  & & & & \\
\end{tabular}
~~~~~~~~~\\
\noindent {\bf Notes to Table:}\\
\noindent The luminosities of the total \otw and \oth line emission are given in columns 2 and 3 for each quasar. \\
\noindent The ratio of nuclear to total luminosity in both \otw and
\oth are given in columns 4 and 5, where the \lq nuclear' luminosity
is measured from the total spectrum extracted from the 19 fibres around the peak fibre. \\
\noindent The position angle of the the optical nebula (column 6) is
 estimated to within $\pm10$\degmark. \\
\noindent $^{\ast}$ Radio jet is deflected to a position angle of
100\degmark\ after 3.6 arcsec. \\
\noindent $^{\dag}$ Position angle of inner jet only; position angle of counter jet-side follows in brackets. \\
\noindent $^{\ddag}$ Optical nebula extends out of the ARGUS aperture.\\
\noindent References for radio maps are given in the final column: (Bo) 1.4GHz  Bogers \etal  1994;
(Br) 4.9GHz Bridle \etal 1994; (Mi) 4.9GHz Miller \etal 1993 \\
\end{table*}

%215 f(o3) 6.2609e-15
%281       9.6316e-15 f(o2) 1.29024446 
%37.43 f(o3) 44.8805771 f(o2) 1.67744327 
%334 fo3 31.362 f(o2) 3.7981801 
%11.72 fo3 55.3652191 fo2 

\section{Summary and Discussion}

\subsection{Morphology}
The quasar nebulosities presented show both smooth (eg 3C323.1) and
clumpy (eg 4C11.72, 4C37.43) distributions, and none are particularly
symmetric about the radio axis. 3C334 and 3C215 both show plumes of
emission, and 3C281 shows only an indistinct elongation to one side.
Nearly all of the quasars are associated with large ($>200$\kpc)
radio sources, and in all but the exception (4C117.2, with a
radio source diameter $<100$\kpc) 
the extent of the optical emission is 
very much smaller (typically around ten per cent; though 
note that both 4C37.43 and 4C11.72 are known to extend
out of the ARGUS aperture at lower flux levels; SM87). Although
estimating a position angle for the main orientation of the nebulae is
subjective (and only good to within $\pm10$\degmark;
Table~\ref{tab:nebprops}), we find a range of behaviour with respect
to the radio source orientation. Both the EELRs of 3C323.1 and 3C281
have their maximum elongation approximately perpendicular to the radio
source axis, that of 4C11.72 is directly coincident with the radio
source structure. The line emission around 4C37.43 shows an indistinct
relation to the radio axis, while the EELR of 3C215 appears to be
completely unrelated (albeit extended perpendicular to the {\em
generic} radio source axis). We do not see any tendency for the EELR
to be aligned with the radio jet in the two jetted radio sources. In
general the distribution of the gas is much less anisotropic than that
around radio-loud quasars at higher redshift where the alignment
effect is observed (Heckman \etal 1991a; Ridgway
\& Stockton 1998).

For several of the systems presented, a large fraction of the total
line luminosity is emitted from the EELR. In two cases (4C11.72 and
4C37.43) about two-thirds of the total \oth is emitted from the EELR;
a similar proportional distribution is seen for the \otw in 4C11.72,
where as one-third of the total [OII] is off-nuclear in 4C37.43. Apart
from 3C323.1 where about a half of the total [OII] and a third of the
total [OIII] are emitted by the EELR, a similar proportional
distribution is seen between the [OII] and [OIII] in the rest of the
quasars (3C334 $\sim65$ per cent; 3C215 $\sim87$ per cent and 3C281
$\sim68$ per cent). It is thus possible that the nebula also emits
along the line of sight to the nucleus, yielding a slight overestimate
to the nuclear contribution to the total line luminosity.

\subsection{Kinematic behaviour}
The EELR studied show a range of kinematic behaviour, three of which
are in some way related to the radio source axes. 3C323.1 displays
possible bipolar rotation $\pm200$\kmps about an axis not quite
aligned with the radio source axis. 
Although the EELR in 4C11.72 also appears at first glance 
to be in general dipolar rotation $\pm200\kmps$ (albeit about an axis
now perpendicular to that of the radio source), there is a region of an
added strong blueshift and possibly of higher ionization where
 coincident with the southern radio lobe. There
is no obvious increase in velocity width in this region, however, that
would be indicative of an interaction between the optical and radio
plasmas. 

4C37.43 shows a strong blueshift with respect to the nucleus
in the two major condensates either side of the quasar, which each
display  a possible dipolar structure
about an axes directed radially out of the nucleus.
This velocity structure precludes the gas we see being in
simple dipolar rotation around the quasar.  
Although the clouds are not wholly coincident with the radio axis, it
is notable that the region of largest blueshift and of higher ionization
in both the East and West clouds, and the highest FWHM in the West
cloud (but not the Eastern cloud) are all where the optical gas is
crosses the radio source axis. Whilst these may be
taken as indicators in favour of a jet-cloud interaction (particularly to the
north-west of the QSO), there is no obvious distortion in the radio
source  to support this interpretation (Miller \etal 1993). 
The off-nuclear clouds may simply be seen in
projection against the quasar; their velocity is the expected order of
magnitude to be contained in a poor cluster of galaxies. Although the
blueshift could alternatively be due to an outflow (eg from the expansion of ionized gas
heated in a cone around the quasar continuum beam) one would
have to appeal to a very distinct geometry or obscuration to explain
the lack of redshifted component. 

The other three quasars appear to have kinematic behaviour that does
not seem to be particularly related to the contained radio source. Even
though \oth emission around 3C281 is not very extended, it is slightly
redshifted (by 300\kmps) with respect to the nucleus. 3C215 and 3C334
both appear to show essentially static systems (all within 150\kmps of
the nucleus). 

Most of the EELR around these quasars show a consistent pattern of
highest line-widths coincident with the quasar nucleus, falling to
typically $\sim200\kmps$ at larger radii.

\subsection{Ionization state}
The ionization state of the emitting regions around each quasar (as
defined by the line ratios between \otwn, \oth and \Hbn) show them to
be very similar to each other (Figs~\ref{fig:allrats} and
\ref{fig:4c37.43bpt}), implying that the ionization of the EELR occurs
within a relatively homogeneous set of physical conditions. The line
ratios and lack of correlation with any extended continuum structure
(SM87) argue against the gas being ionized {\em in situ} by hot stars.
Given the proximity to a luminous quasar nucleus, one can assume that
illumination by the UV continuum of the active nucleus provides the
{\em minimum} source of ionization. The fact that the morphology of
the gas is not in all cases strongly aligned with the radio source
axis implies that the quasar UV radiation is not strongly confined
(either by beaming or shielding) to escape only along this axis.

The only regions of higher ionization observed are those directly
coincident with the radio source axis, most notable to the eastern side
of 4C37.43 and more marginally to the south-east of 4C11.72. Unlike
the case of 3C254 (Crawford \& Vanderriest 1997) there is no {\em
clear} evidence for a jet-cloud interaction to be occurring in these
regions (\ie the {\em combination} of increased line-width, distorted
radio structure, a large velocity gradient). We note, however, that
any shocks caused by compression where the EELR is spatially
coincident with the radio plasma in these regions will only add to the
default photo-ionization provided by the quasar.

\subsection{Pressure Estimates}
Assuming that the dominant source of ionization for the emitting gas
is the UV continuum of the quasar nucleus, we can use the line
intensity ratios observed in the EELRs to infer physical properties
of the emitting gas. We match the observed ratios to those obtained
from photoionization modelling using CLOUDY 90.04 (Ferland 1996),
assuming the nuclear continuum is not very different from that we see,
and that the gas lies at the projected separation from the nucleus
(see \eg Forbes \etal 1990; Crawford \etal 1991 for a more detailed
discussion of this technique and the assumptions thereof).

We estimate the shape of the ionizing continuum from archival ultraviolet
HST and X-ray ROSAT PSPC spectral data of the quasars. We first fit
the 1050-3000\AA\ HST spectrum (corrected for Galactic reddening) by a
power-law of slope $\alpha_{UV}$, where
$f_\nu\propto\nu^{\alpha_{UV}}$. [If defining the continuum level is
complicated by the presence of broad emission bands and lines, we
fit the spectrum using only the continuum bands listed by Zheng
\etal (1997)]. Similarly we fit the 0.2--2\keV\ ROSAT PSPC spectrum by a
power-law slope of $\alpha_X$; none of the quasars in this sample
required any significant absorption above that expected from the
line-of-sight Galactic column density. There is a problem, however, in
that the ROSAT PSPC consistently measures power-law slopes that are
{\em steeper} by $\Delta\alpha_X\sim0.5$ than is derived from
observations of the same objects by other X-ray missions (Fiore \etal
1994; Laor \etal 1997; Iwasawa, Fabian \& Nandra 1999). We correct for
this effect by assuming X-ray slopes shallower by
$\Delta\alpha_X\sim0.5$ than we fit from the PSPC spectra. For the two
quasars with no PSPC observation (4C11.72 and 3C281) we assume the
X-ray continuum to follow the canonical power-law slope for a
radio-loud quasar of $\alpha_{X}=-0.5$ (Schartel \etal 1996; Laor
\etal 1997 and taking into account the artificially steep slopes that
ROSAT measures), and fix the relative normalization between the UV and
X-ray bands from the ROSAT X-ray flux given by Brinkmann \etal (1995).

We thus represent the nuclear continuum over the UV/X bandwidths by
power-laws of slopes $\alpha_{UV}$ and $\alpha_X$, with the relative
normalization fixed so that $\alpha_{OX}$ (the slope between 2500\AA\
-- 2\keV; see eg Zamorani \etal 1981) fits the observed fluxes. The
whole spectrum is normalized to the observed luminosity of the quasar.
The [OIII]/\Hb ratio observed in the EELR also allows us to constrain
the high-energy end to any `Blue Bump' component, parameterized by the
temperature $T$ of an exponential cutoff to the lower-energy continuum
of slope $\alpha_{UV}$. The values for these parameters defining the
input AGN continuum are listed for each quasar in
Table~\ref{tab:ioncntm}. The situation is complicated for
4C37.43, where the higher ionization regions in some parts of the EELR
(Fig~\ref{fig:4c37.43bpt}) would require a different ionizing
continuum than do the ratios from the remainder of the diffuse emission. It is
probable that the higher ionization is caused by some extra source of
ionization {\em in situ} in these regions, so we do not attempt to
model these regions in detail. 
%We do note, however, that any extra
%ionization will only serve to increase the pressure estimated in this
%higher ionization region. 

\begin{table*}
\caption{Nuclear ionizing continuum \label{tab:ioncntm} }
\begin{tabular}{lcccr}
Name    & $\alpha_{UV}$ & $\alpha_{X}$ & $\alpha_{OX}$ & $T$ ($10^5$K) \\
\hline 
3C323.1 & --0.6 & --1   & --1.4 &  2.5 \\
4C11.72 & --0.9 & --0.5 & --1.6 & 10 \\
4C37.43 &  0   & --0.7 & --1.4 &  2.25\\
3C215   & --0.5 & --0.5 & --1.1 & $>$3.5$^\ast$ \\
3C334   & --0.5 & --0.8 & --1.45 & 6.5  \\
3C281   & --1   & --0.5 & --1.2 &  $>$3$^\ast$ \\
& & & & \\
\end{tabular}
~~~~~ \newline
\noindent $\alpha$ for all bands is defined as $f_\nu\propto
\nu^{\alpha}$. \\
\noindent $\alpha_X$ has been corrected for the over-steepening of
PSPC spectra of $\Delta\alpha_X\sim0.5$, as mentioned in the text. \\
\noindent $^\ast$ $T$ can only be constrained to be a lower 
limit where only a lower limit to the [OIII]/\Hb ratio 
can be measured. \\
\end{table*}

\begin{figure}
% /home/csc/text/papers/qso/newpress.results
\psfig{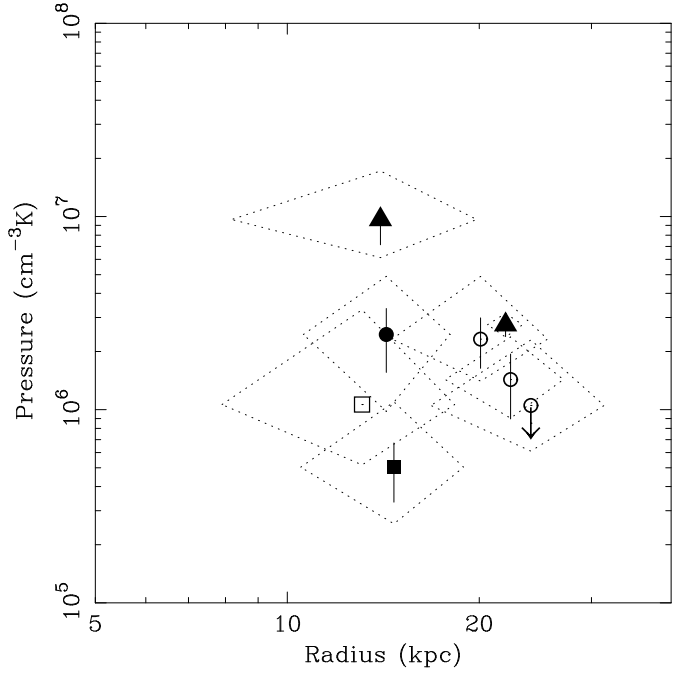}
\caption{\label{fig:allpress}
Pressures deduced from the CLOUDY runs as a function of radius from
the quasar for all objects except 4C37.43 (for which see
Fig~\ref{fig:4c37.43press}). 3C323.1 is marked by a solid circle,
4C11.72 by open circles, the limits to 3C215 by an open square, 3C334
by solid triangles and 3C281 by a solid square. The vertical error
bars only encompass the errors on [OIII]/[OII] caused by the
$\Delta\chi^2$=1 confidence values from the fits to the emission-line
intensities. The larger error diamonds around each point mark the
range of radius from which the line ratios have been extracted, and
also the larger variation in pressure deduced if the full variation in
radius is taken into account. Note that the point marked for 3C281
represents a lower limit to the derived pressure, due to our measuring
only a lower limit to the [OIII]/\Hb ratio in the extended gas. This
also applies for the point for 3C215; however, as we also obtain only
an upper limit to [OIII]/[OII] the pressure inferred could also be
less. }\end{figure}

\begin{figure}
% fig is in /data/csc/argus/jundata/4c37.43/newcloudy/press.out
\psfig{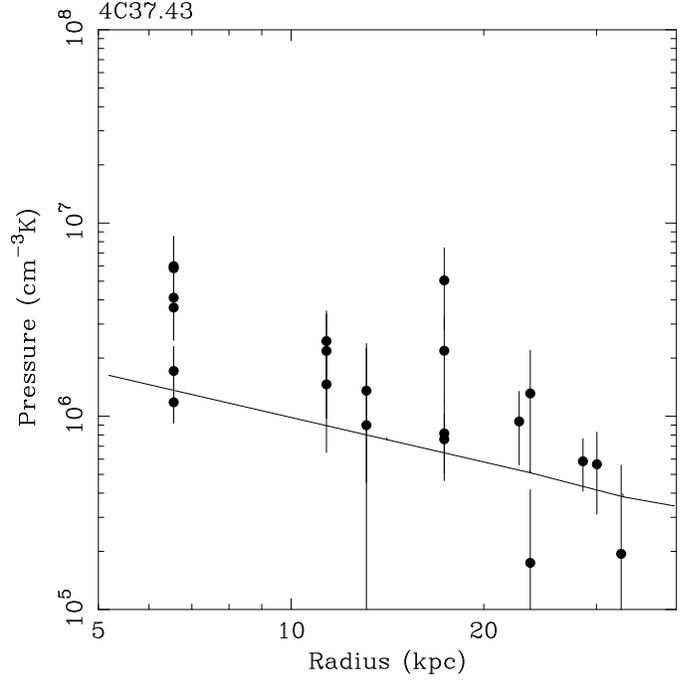}
\caption{\label{fig:4c37.43press}
Radial pressure profile deduced from the emission line nebula around
4C37.43, excluding the higher-ionization blobs. The pressures are deduced by 
photo-ionization calculations in CLOUDY as described in the text,
modelling the blue bump with a temperature of 
$2.25\times10^5$\K. For comparison, the solid line shows the pressure profile within
the intracluster medium around the radio source PKS1404-267 derived
from X-ray observations taken with ROSAT (from Johnstone \etal 1998). 
}\end{figure}

The quasar continuum is assumed incident on a neutral gas cloud at the
projected radius of the EELR from which the [OII]/[OIII] ratios shown
in Table~\ref{tab:linerats} and Figs~\ref{fig:allrats} and
\ref{fig:4c37.43bpt} are extracted. The cloud parameters are varied in
CLOUDY until the runs predict a good match to the observed line
ratios. The pressures deduced within the EELR of all six
quasars show a very similar and high range of values, all over
$10^5$\pcmcuK\ for projected radii of 15-25\kpc\ from the nucleus
(Figs~\ref{fig:allpress} and
\ref{fig:4c37.43press}).
Uncertainties in these estimates are due to
the errors in the [OII]/[OIII] ratio from the emission-line fitting,
typically around $\pm$35 per cent; uncertainties in the temperature of
the `Blue Bump', where the errors in [OIII]/\Hb could change the bump
temperature and thus the inferred pressure by typically $\pm$20 per
cent. The main potential source for error is the range in radius over
which the line ratios are averaged; at the inner (outer) part of this
range the same line ratio would match a gas pressure typically a
factor of two greater (smaller) than that inferred. This error margin
is shown by the error diamonds in the Fig~\ref{fig:allpress}, where
the error bars represent the errors due to the uncertainty only in the
measurement of the [OIII]/[OII] line ratio. Note though, that the
lower limit to the temperature of the bump in 3C281 and 3C215 implies
that the pressures inferred could also be greater at a given distance.

\subsection{Interpretation of the immediate environment}
The pressures inferred within the extended emission-line regions
around the six quasars are all compatible with those measured in the
intracluster medium of nearby clusters of galaxies. Furthermore, the
range of pressures deduced are consistent with a moderate
cooling flow (of tens to a couple of hundred solar masses a year)
operating in the nearby environment of the quasar. The radial pressure
profile of the ionized gas around 4C37.43 has a proportionality of
$P\propto r^{-0.8\pm0.2}$ (Fig~\ref{fig:4c37.43press}). This is
compatible with the EELR being in pressure equilibrium with hot gas
following the $\sim r^{-1}$ distribution expected for distributed mass
deposition in an isothermal halo. For comparison, in
Fig~\ref{fig:4c37.43press} we also plot the pressure profile derived
from a deprojection of ROSAT HRI X-ray data of A3581
from Johnstone, Fabian \& Taylor (1998). A3581 is an interesting
analogy to 4C37.43, as it is a poor cluster, hosting the FR~II radio
source PKS1404-267 at its core. The X-ray data show a moderate cooling
flow with a mass deposition rate of $\sim80$\Msunpyr to be occurring
within the intracluster medium. The similarity of its pressure profile to our
optically-deduced ones supports the inference of a cooling flow around
these quasars, despite the fact that only three (3C215, 3C281 and
3C323.1) lie in environments that can be classified from field optical
galaxy counts as richer than Abell class 0 (Ellingson \etal 1991; Yee
\& Green 1984, 1987; 3C334 and 4C11.72 lie less clustered regions,
and 4C37.43 lies in a comparatively sparse environment.)
In a hierarchical model for the evolution of clusters, it may well be
that quasars -- and their relatively undisturbed cooling
flows -- can only survive to lower redshifts in the {\em poorer}
clusters that are sufficiently isolated to have so far escaped merging
with richer systems (see e.g. Fabian \& Crawford 1990). 

Our interpretation is also supported by the recent direct detections
of the X-ray emission from the host clusters of 3C215, 3C281 and 3C334
(among a handful of such detections; Crawford \etal 1999; Hardcastle
\& Worrall 1999) support this suggestion. In these cases, the extended
X-ray component is consistent with thermal emission from the
intracluster medium of moderately rich host clusters to the quasars,
with inner regions dense enough to be part of a cooling flow.

%  We find no
%obvious trend between the optical quasar-galaxy covariance function
%($B_{gq}$) and the pressure we deduce within the EELR.

%4C37.43 also presents a problem in that the pressure deduced
%(especially in the high-ionization regions) is very close to the
%'maximal pressure' where the cooling time of the gas would equal its
%free-fall time (Bremer etal 1992). Clearly a cooling flow with a mass
%deposition rate of thousands of solar masses a year is not occurring.
%%%%%%%%%%%HELP %%%%%%%%%%%%%%

\subsection{Relation of the EELR to the radio source}

Five out of six of our quasars are associated with radio sources that
are over 200\kpc\ in diameter (Table~\ref{tab:nebprops}). Amongst
these are three with settled or quiescent kinematic behaviour
(3C323.1, 3C334 and 3C215), and one with a little (3C281) and one with
a large (4C37.43) amount of velocity structure. The quasar with the
smallest contained radio source, 4C11.72, not surprisingly has the
most kinematically active EELR.
Naively this mostly conforms to the similar division of behaviour seen
in the emission-line nebulae of higher-redshift radio galaxies by Best
\etal 1999, whereby the radio sources with smaller linear sizes
($\le150\kpc$) show clear indications of physical interactions between
the radio and optical plasmas, in terms of kinematic behaviour,
ionization state and flux. Larger systems have EELRs more consistent
with photo-ionization by an embedded quasar. We cannot test for the
role of shock ionization in our sources as the diagnostic UV lines are
not observable at the redshifts of our sample; \oth and \otw do not
allow us to clearly discriminate between these possibilities. However, we note that any
extra sources of ionization (such as shocks from jet-cloud
interactions) will act to exacerbate the high pressures inferred from
the low ionization state of the EELR. 
The exception to this model is the EELR around 4C37.43, which shows
clear regions of higher ionization aligned along the radio axis.
Although only associated with an increased FWHM to the north-west side
of the quasar, it is possible that extra ionization in the form of
shocks may produce this ionization differential within the nebula.

%\section{Other physical properties of the EELR}
%*** {\em NOT SURE HOW TO PROGRESS HERE, given some doubt about how truthful
%the CLOUDY findings are and thus how appropriate further extrapolation
%is ***. }
%[(notes for 3C323.1 only so far:)
%By comparison of the observed fluxes to those predicted from a typical
%CLOUDY run that matches the line ratios, we find that the covering
%fraction of an individual shell one fibre thick is around 0.5 per
%cent. Given there are up to nine such layers within the nebula to the
%south-east of the nucleus, we infer a total covering fraction of
%$\sim5$ per cent. The CLOUDY results show that the [OIII] and [OII]
%line emission is emitted within a column density of $10^{20}$\pcmsq.
%Assuming a gas temperature of $10^4$K, and a pressure of
%$\sim2.6\times10^6$\pcmcuK, the ratio of the depth to the observed
%scale of the nebula gives an estimate of the volume filling factor as
%$\sim3\times10^{-7}$. Approximating the whole EELR as a cylinder
%14\kpc\ wide and 28\kpc\ long at an average pressure of
%$\sim2.6\times10^6$\pcmcuK, we estimate the total mass of the visible
%ionized gas to be $9\times10^6$\Msun.]
%

\section{Conclusions}

We have presented two-dimensional maps of the distribution, velocity
structure and ionization state of the extended emission-line regions
around six intermediate-redshift quasars. The EELRs show a diverse
behaviour in morphology -- including relation to the radio source axis
-- and kinematics. Despite this, they all have surprisingly similar 
-- and low -- 
ionization, as measured by the ratio of \oth and \otwn. We thus infer
a confining medium around all the quasars to sustain the EELR at such
low ionization in the proximity of a strong ionizing quasar nucleus. 

\section{Acknowledgments}
 CSC thanks both the PPARC and the Royal Society for financial support
 while this work was in progress. This research has made use of the
 NASA/IPAC Extragalactic Database (NED) which is operated by the Jet
 Propulsion Laboratory, California Institute of Technology, under
 contract with the National Aeronautics and Space Administration. We
 have also made use of data obtained from the Leicester Database and
 Archive Service at the Department of Physics and Astronomy, Leicester
 University.


\begin{thebibliography}{}
\bibitem[]{} Bahcall J. N., Kirhakos, S.,  Saxe  D. H., Schneider D. P., 1997, ApJ, 479, 642
%\bibitem[]{} Barthel P. D., 1989, ApJ, 336, 606
\bibitem[]{} Best P., Longair M. S., Rottgering H. J. A., 1998, MNRAS, 295, 549
\bibitem[]{} Best P., Rottgering H. J. A., Longair M. S., 1999, MNRAS,
in press (astro-ph/9908211)
\bibitem[]{} Block D. L., Stockton A., 1991, AJ, 102, 1928 
\bibitem[]{} Bogers W. J., Hes R., Barthel P. D., Zensus J. A., 1994, A\&AS, 105, 91
\bibitem[]{} Bohlin R. C., Savage B. D., Drake J. F., 1978, ApJ, 224, 132
\bibitem[]{} Boisson C., Durret F., Bergeron J., Petitjean P. , 1994, A\&A, 285, 377
\bibitem[]{} Bower R., Smail I., 1997, MNRAS, 290, 292
\bibitem[]{} Bremer M. N., Crawford C. S., Fabian A. C., Johnstone R. M., 1992, MNRAS, 254, 614
\bibitem[]{} Bridle A. H., Hough D. H., Lonsdale C. J., Burns J. P., Laing R. A., 1994, AJ, 108, 766 
\bibitem[]{} Brinkmann W., Siebert J., Reich W., Furst E., Reich P.
             Voges W., Trumper J., Wielebinski R., 1995, A\&A Suppl, 113, 347
\bibitem[]{} Brotherton M. S., 1996, ApJS, 102, 1
\bibitem[]{} Canalizo G., Stockton A., 1997, ApJLett, 480, L5
\bibitem[]{} Cardelli J. A., Clayton G. C., Mathis J. S., 1989, ApJ, 345, 245
\bibitem[]{} Carilli C. L., Owen F. N., Harris D. E., 1994, AJ, 107, 480 
\bibitem[]{} Carilli C. L. et al, 1997, ApJS, 109, 1 
\bibitem[]{} Chatzichristou E.T., Vanderriest C., Jaffe W., 1999, A\&A, 343, 407
\bibitem[]{} Crawford C. S., Fabian A. C., 1989, MNRAS, 239, 219
\bibitem[]{} Crawford C. S., Fabian A. C., 1993, MNRAS, 260, L15
\bibitem[]{} Crawford C. S., Fabian A. C., 1995, MNRAS, 273, 827
\bibitem[]{} Crawford C. S., Fabian A. C., 1996a, MNRAS, 281, L5
\bibitem[]{} Crawford C. S., Fabian A. C., 1996b, MNRAS, 282, 1483
\bibitem[]{} Crawford C. S., Fabian A. C., Johnstone R. M., 1988, 235, 183
\bibitem[]{} Crawford C. S., Lehman I., Fabian A. C., Bremer M. N.,
Hasinger G., 1999, MNRAS, in press (astro-ph/9904371)
\bibitem[]{} Crawford C. S., Fabian A. C., George I. M., Naylor T. N., 1991, MNRAS, 248, 139 
\bibitem[]{} Crawford C. S., Vanderriest C., 1996,  MNRAS, 283, 1003
\bibitem[]{} Crawford C. S., Vanderriest C., 1997,  MNRAS, 285, 580
\bibitem[]{} Deltorn J. -M., Le Fevre O., Crampton D., Dickinson M., 1997, ApJ, 483, L21
\bibitem[]{} Dickinson M.,  1997 {\sl HST and the High Redshift
Universe}, eds, Tanvir N., Aragon-Salamanca A.,  Wall J.V., 
published by World Scientific.
\bibitem[]{} Dickinson \etal 1999, submitted to ApJ 
\bibitem[]{} Durret F., Pecontal E., Petitjean P., Bergeron J., 1994, A\&A, 291, 392
\bibitem[]{} Ellingson E., Yee H. K. C., Bechtold J., Dobrzycki A., 1994,
             AJ, 107, 1219
\bibitem[]{} Ellingson E., Yee H. K. C., Green R. F. 1991, ApJ, 371 49 
\bibitem[]{} Fabian A. C., 1994, ARAA, 32, 277
\bibitem[]{} Fabian A. C., Crawford C. S., 1990, MNRAS, 247, 439
\bibitem[]{} Ferland G. J., 1996, {\em Hazy, a Brief Introduction to
            Cloudy}, University of Kentucky Department of
            Physics and Astronomy Internal Report, University of Kentucky
\bibitem[]{} Fiore F.,  Elvis M., McDowell J. C., Siemiginowska A., Wilkes B.  J., 1994, ApJ, 431, 515 
\bibitem[]{} Forbes D. A., Crawford C. S., Fabian A. C., Johnstone R. M., 1990, MNRAS, 244, 680
\bibitem[]{} Gunn J., 1971, ApJLett, 164, L113
\bibitem[]{} Garrington S. T., Conway, R. G., 1991. MNRAS, 250, 198
\bibitem[]{} Hardcastle M., Worrall D. M., 1999, MNRAS in press, (astro-ph/9907034)
\bibitem[]{} Heckman T. M., Lehnert M. D., van Breugel W. J. M., Miley G. K., 1991a, ApJ 370, 78
\bibitem[]{} Heckman T. M., Lehnert M. D., Miley G. K., van Breugel W. J. M., 1991b, ApJ 381, 373 
\bibitem[]{} Hes R., 1995, Ph.D. Thesis, University of Groningen
\bibitem[]{} Hes R., Barthel P. D., Fosbury R. A. E., 1996, A\&A, 313, 423 
\bibitem[]{} Hickson P.,  Hutchings J. B., 1987, ApJ, 312, 518
\bibitem[]{} Hill G. J., Lilly S. J., 1991, ApJ,  367, 1
\bibitem[]{} Hintzen P., 1984, ApJSuppl, 55, 533
\bibitem[]{} Hintzen P., Scott J. S., 1978, ApJL, 224, L47
\bibitem[]{} Hutchings J. B., 1992, AJ, 104, 1311
\bibitem[]{} Hutchings J. B., Crampton D., 1990, AJ, 99, 37
\bibitem[]{} Hutchings  J. B., Johnson I., Pyke R., 1988, ApJSuppl, 66, 361
\bibitem[]{} Hutchings J. B., Neff S. G., 1992, AJ, 104, 1
\bibitem[]{} Iwasawa K., Fabian A. C., Nandra K., 1999, MNRAS, 307, 611
\bibitem[]{} Johnstone R.M., Fabian A. C., \& Taylor G. B., 1998, MNRAS, 298, 854
\bibitem[]{} Laor A., Fiore F., Elvis M. J., Wilkes B. J., McDowell J. C., 1997, ApJ, 477, 93
\bibitem[]{} Lawrence C. R. 1990, {\em Parsec Scale Radio Jets}, eds
             Zensus J. A. \& Pearson T. J., CUP, p280. 
\bibitem[]{} McLeod, K. K., Rieke G. H.,  1994, ApJ, 431, 137
\bibitem[]{} Miller P., Rawlings S., Saunders R., 1993, MNRAS, 263, 425 
\bibitem[]{} Neugebauer G., Matthews K., Armus L., 1995, ApJLett, 455, L123 
\bibitem[]{} Oemler A., Gunn J. E., Oke J. B., 1972, ApJL, 176, L47
\bibitem[]{} Ridgway S. E., Stockton A., 1998, AJ, 114, 511
\bibitem[]{} Robinson L. B., Wampler E. J., 1972, ApJ, 171, L83
\bibitem[]{} Schartel N. \etal, 1996, MNRAS, 283, 1015
\bibitem[]{} Stark A. A., Gammie C. F., Wilson R. W., Bally J., Linke R. A., Heiles C., Hurwitz M.,
              1992, ApJS, 79, 77
\bibitem[]{} Stockton A., Ridgway S. E., 1991, AJ, 102, 488 
\bibitem[]{} Stockton A., MacKenty J. W., 1987, ApJ, 316, 584 (SM87)
\bibitem[]{} Tennant A. F., 1991,  {\it NASA  Technical Memorandum}, 4301
\bibitem[]{} Vanderriest C., 1995, ASP Conf. Ser., 71, 209
\bibitem[]{} Worrall D. M., Lawrence C. R., Pearson T. J., Readhead A. C. S., 1994, ApJ, 420, L17
%\bibitem[]{} Worrall D. M., Tananbaum H., Giommi P., Zamorani G., 1987, ApJ, 313, 596 
\bibitem[]{} Yates M. G., Miller L., Peacock J. A., 1989, MNRAS 240, 129
\bibitem[]{} Yee H. K. C., Green R. F., 1984, ApJ, 280, 79
\bibitem[]{} Yee H. K. C., Green R. F., 1987, ApJ, 319, 28
\bibitem[]{} Zamorani G. \etal, 1981, ApJ, 245, 357 
\bibitem[]{} Zheng W., Kriss G. A., Telfer R. C., Grimes J. P.,
             Davidsen A. F., 1997, ApJ, 475, 469 
\end{thebibliography}
\end{document}